\documentclass[aps,pra,twocolumn,showpacs,notitlepage,superscriptaddress,letterpaper]{revtex4-1}
\usepackage{graphicx, amsmath, amssymb, amsfonts, pifont, bm, cancel}
\usepackage[usenames]{color}
\usepackage{subfigure}
\usepackage[normalem]{ulem} 
\usepackage{circuitikz}
\usepackage{hyperref}
\usepackage{upgreek}

\definecolor{MyDarkBlue}{rgb}{0,0,1}

\newcommand{\beq}{\begin{equation}}
\newcommand{\eeq}{\end{equation}}
\newcommand{\bel}{\begin{align*}}
\newcommand{\tamam}{\end{align*}}

\newcommand{\ket}[1]{|#1\rangle}

\newcommand{\beqa}{\begin{eqnarray}}             
\newcommand{\eeqa}{\end{eqnarray}}               
\newcommand{\bra}[1]{\langle#1\vert}                 

\newcommand{\taudelay}{\tau_{\text{d}}}
\newcommand{\omegamod}{\omega_{\text{mod}}}
\newcommand{\omegage}{\omega_{ge}}
\newcommand{\omegageint}{\omegage^{\prime}}
\newcommand{\omegaef}{\omega_{fe}}
\newcommand{\omegares}{\omega_{r}}
\newcommand{\gvacres}{g_{r}}
\newcommand{\gvacMMuc}{g_{\text{uc}}}
\newcommand{\omegaMMres}{\omega_{0}}
\newcommand{\annharm}{\eta}
\newcommand{\dbar}{\bar{d}}
\newcommand{\GammaoneD}{\Gamma_{\text{1D}}}
\newcommand{\Besselone}{\mathcal{J}_{1}}
\newcommand{\coupcap}{C_g}
\newcommand{\gndcap}{C_0}
\newcommand{\meanind}{L_0}
\newcommand{\qubitcap}{C_q}
\newcommand{\fiftyOhm}{$50$-$\Omega$}
\newcommand{\raisingk}{\hat{a}_k^\dagger}
\newcommand{\loweringk}{\hat{a}_k}
\newcommand{\loweringq}{\hat{\sigma}^-}
\newcommand{\raisingq}{\hat{\sigma}^+}
\newcommand{\dk}{\text{d}k}
\newcommand{\taugroup}{\tau_{g}}
\newcommand{\OmegaWG}{\Omega_{\text{WG}}} 
\newcommand{\Tf}{T_{\text{f}}}
\newcommand{\Qone}{$\text{Q}_1$}
\usepackage{physics}

\begin{document}

\title{Collapse and Revival of an Artificial Atom Coupled to a Structured Photonic Reservoir}

\author{Vinicius~S.~Ferreira}
\thanks{These two authors contributed equally}
\author{Jash~Banker}
\thanks{These two authors contributed equally}
\author{Alp~Sipahigil}
\author{Matthew~H.~Matheny}
\author{Andrew~J.~Keller}
\author{Eunjong~Kim}
\author{Mohammad~Mirhosseini}
\affiliation{Kavli Nanoscience Institute and Thomas J. Watson, Sr., Laboratory of Applied Physics, California Institute of Technology, Pasadena, California 91125, USA.}
\affiliation{Institute for Quantum Information and Matter, California Institute of Technology, Pasadena, California 91125, USA.}
 
\author{Oskar~Painter}
\email{opainter@caltech.edu}
\homepage{http://copilot.caltech.edu}
\affiliation{Kavli Nanoscience Institute and Thomas J. Watson, Sr., Laboratory of Applied Physics, California Institute of Technology, Pasadena, California 91125, USA.}
\affiliation{Institute for Quantum Information and Matter, California Institute of Technology, Pasadena, California 91125, USA.}

\date{\today}

\begin{abstract}
{A structured electromagnetic reservoir can result in novel dynamics of quantum emitters. In particular, the reservoir can be tailored to have a memory of past interactions with emitters, in contrast to memory-less Markovian dynamics of typical open systems. In this Article, we investigate the non-Markovian dynamics of a superconducting qubit strongly coupled to a superconducting slow-light waveguide reservoir.  Tuning the qubit into the spectral vicinity of the passband of this waveguide, we find non-exponential energy relaxation as well as substantial changes to the qubit emission rate. Further, upon addition of a reflective boundary to one end of the waveguide, we observe revivals in the qubit population on a timescale $30$ times longer than the inverse of the qubit's emission rate, corresponding to the round-trip travel time of an emitted photon. By tuning of the qubit-waveguide interaction strength, we probe a crossover between Markovian and non-Markovian qubit emission dynamics. These attributes allow for future studies of multi-qubit circuits coupled to structured reservoirs, in addition to constituting the necessary resources for generation of multi-photon highly entangled states.
}
\end{abstract}
\maketitle

\clearpage

\section{Introduction}
\label{intro}
Spontaneous emission by a quantum emitter into the fluctuating electromagnetic vacuum is an emblematic example of Markovian dynamics of an open quantum system \cite{weisskopf1930berechnung}. However, modification of the electromagnetic reservoir can drastically alter this dynamic, introducing ``non-Markovian" memory effects to the emission process, a consequence of information back-flow from the reservoir to the emitter~\cite{haase2018controllable,hoeppe2012direct,liu2011experimental,madsen2011observation}. There have been several studies investigating non-Markovian effects on the preservation of quantum information and multipartite entanglement~\cite{gonzalez2013non,bellomo2007non}. These studies have generated interest in leveraging long-lived environmental correlations for stabilization and synthesis of many-body, arbitrary quantum states of a quantum system~\cite{huelga2012non,reich2015exploiting,cheng2016preservation,bylicka2014non, pichler2017universal}.

Studies of non-Markovian physics are readily achieved by strongly coupling an emitter to a single-mode waveguide -- a one-dimensional (1D) reservoir with a continuum of states.  Waveguides which break translational symmetry, or which host resonant elements within the waveguide, are of particular interest in this regard owing to the structure in their spectrum~\cite{tan2008chip,caloz2011metamaterial,saynatjoki2007dispersion}.  For example, rich phenomena emerge upon constriction of the 1D continuum of guided modes to a transmission band of finite bandwidth, with sharp transitions in the photonic density of states (DOS) occurring at the bandedges. These phenomena include non-exponential radiative decay, finite light trapping close to the bandedge, and the formation of protected atom-photon bound states far from the continuum \cite{john1994spontaneous,shen2019non,gonzalez2017markovian, lambropoulos2000fundamental, vats1998non,bellomo2008entanglement}. Spectral constriction of the continuum, and the concomitant frequency dispersion, can result in the slowing of light propagation which enables observation of additional non-Markovian phenomena. For instance, by placing a reflective boundary on one end of a slow-light waveguide, i.e. a mirror, a fraction of the emitter's radiation is reflected back from the mirror, thus inducing energy back-flow from the waveguide reservoir at significantly delayed timescales~\cite{Tufarelli2014, pichler2016photonic,hoi2015probing}. Surprisingly, this deceptively simple mechanism of non-Markovian time-delayed feedback can allow for generation of multi-dimensional photonic cluster states by a single emitter, provided that $\taudelay\GammaoneD \gg 1$, where $\GammaoneD$ is the emitter's emission rate into the waveguide and $\taudelay$ is the round-trip travel time of an emitted photon~\cite{pichler2017universal}. 



Superconducting microwave circuits incorporating Josephson-Junction-based qubits~\cite{Schoelkopf2008,Devoret2013} represent a near-ideal test bed for studying the quantum dynamics of emitters interacting with a 1D continuum~\cite{van2013photon,lalumiere2013input}.  In comparison to solid-state and atomic optical systems~\cite{vetsch2010optical,yu2014nanowire,javadi2015single,bhaskar2017quantum}, superconducting microwave circuits can be created at a deep-sub-wavelength scale, giving rise to strong qubit-waveguide coupling far exceeding other qubit dissipative channels.  This has enabled a variety of pioneering experiments probing qubit-waveguide radiative dynamics, employing waveguide spectroscopy~\cite{sundaresan2019interacting,liu2017quantum, hoi2015probing, andersson2019non}, time-dependent qubit measurements~\cite{mirhosseini2018superconducting, mirhosseini2019cavity, zhong2019violating, bienfait2019phonon} and analysis of higher-order field correlations~\cite{hoi2012generation, eichler2012observation}. Recent experiments have also explored the coupling of superconducting qubits to acoustic wave devices, demonstrating the capability of these systems to produce significant time-delayed feedback~\cite{andersson2019non, bienfait2019phonon}, albeit with other challenges such as acoustic wave back-scattering, limited acousto-electric coupling, and quantum-limited detection of acoustic fields. 

In this work we develop an all-electrical slow-light waveguide consisting of a superconducting metamaterial waveguide with a highly structured 1D continuum, resulting in significant retardation of propagating microwave fields over a broad bandwidth.  By strongly coupling Xmon-style superconducting qubits~\cite{Koch2007,Barends2013} to the slow-light waveguide, we explore through both spectroscopic field measurements and time-dependent qubit measurements, the properties of this system deep within the non-Markovian limit.  By terminating one-end of the slow-light waveguide with a reflective boundary, we also study the time-delayed feedback of emitted light pulses from the qubit (achieving $\taudelay\GammaoneD \approx 30$), providing insight into the attainable fidelity and scale of the aforementioned multipartite entanglement proposal~\cite{pichler2017universal} in such a physical system.        

\section{Slow-Light Metamaterial Waveguide}
\label{MM_wqveguide}
In prior work studying superconducting qubit emission into a photonic bandgap waveguide~\cite{mirhosseini2018superconducting}, we employed a metamaterial consisting of a coplanar waveguide (CPW) periodically loaded by lumped-element resonators.  In that geometry, whose circuit model simplifies to a transmission line with resonator loading in parallel to the line, one obtains high efficiency transmission with a characteristic impedance approximately that of the standard CPW away from the resonance frequency of the loading resonators, and a transmission stopband near resonance of the resonators.  In contrast, here we seek a waveguide with high transmission efficiency and slow-light propagation within a transmission passband.  In addition to a metamaterial design that optimizes the slow-light delay for a given chip area, secondary considerations include a modular design that can be reliably replicated at the unit cell level without introducing spurious cell-to-cell couplings, and a method for matching to external input and output lines that avoids unintended reflections and resonances within the transmission passband.               

Large delay per unit area can be obtained by employing a network of sub-wavelength resonators, with light propagation corresponding to hopping from resonator-to-resonator at a rate set by near-field inter-resonator coupling.  This area-efficient approach to achieving large delays is well-suited to applications where only limited bandwidths are necessary.  In optical photonics applications, this sort of scheme has been realized in what are called coupled-resonator optical waveguides, or CROW waveguides~\cite{yariv1999coupled,Notomi2008}.  Here we employ a periodic array of capacitively coupled, lumped-element microwave resonators to form the waveguide. Such a resonator-based waveguide supports a photonic channel through which light can propagate, henceforth referred to as the passband, with bandwidth approximately equal to four times the coupling between the resonators, $J$. The limited bandwidth directly translates into large propagation delays; as can be shown (see App.~\ref{App:Theory}), the delay in the resonator array is roughly $\omegaMMres/J$ longer than that of a conventional CPW of similar area, where $\omegaMMres$ is the resonance frequency of the resonators.   

\begin{figure}[tbp]
\centering
\includegraphics[width = \columnwidth]{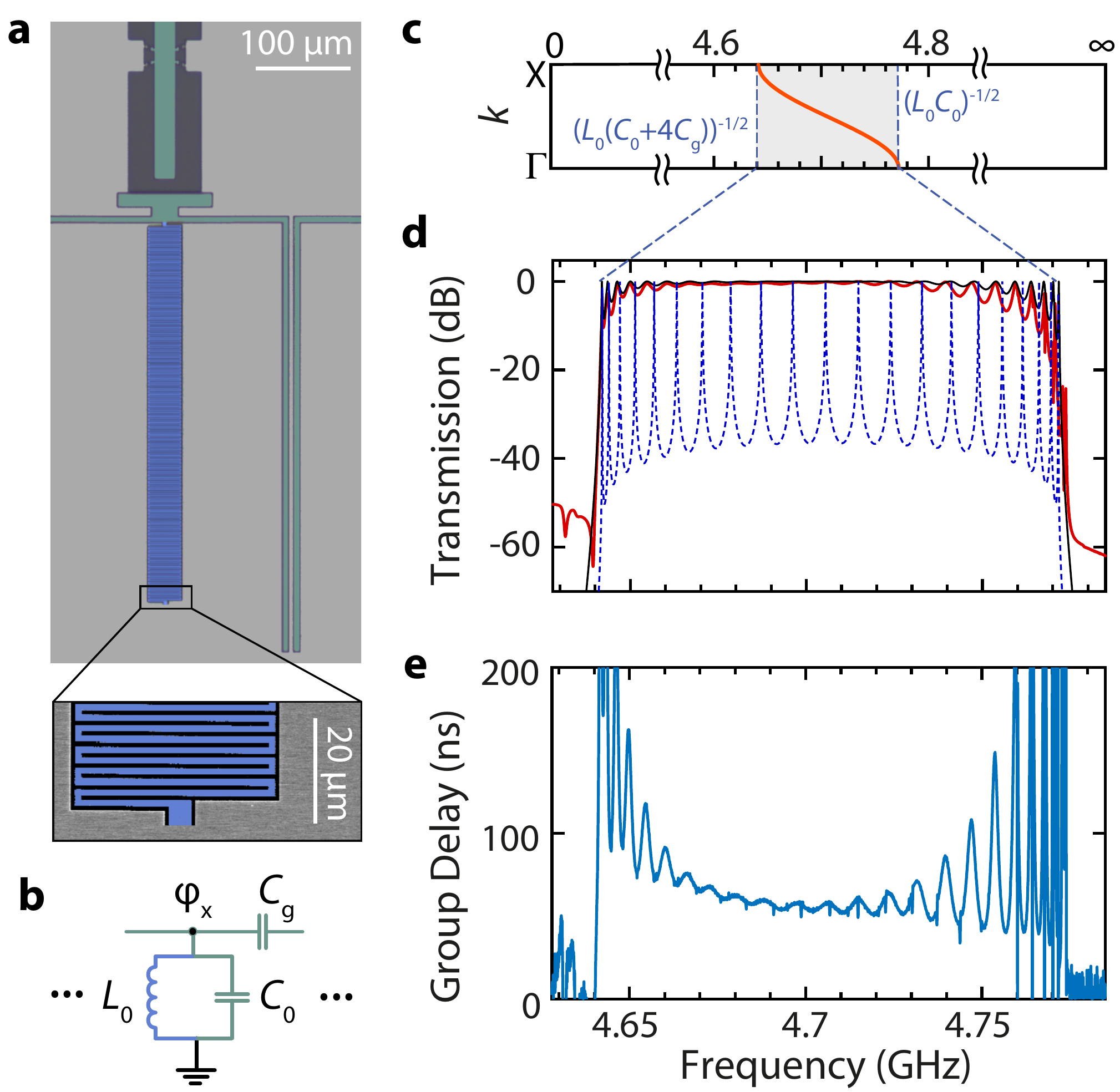}
\caption{\textbf{Microwave Coupled Resonator Array Slow-light Waveguide.} \textbf{a}, Optical image of a fabricated microwave resonator unit cell. The capacitive elements of the resonator are false colored in green, while the inductive meander is false colored in blue. The inset shows a false colored SEM image of the bottom of the meander inductor, where it is shorted to ground. \textbf{b}, Circuit diagram of the unit cell of the periodic resonator array waveguide. \textbf{c}, Theoretical dispersion relation of the periodic resonator array. See App.~\ref{App:Theory} for derivation. \textbf{d}, Transmission through a metamaterial slow-light waveguide spanning $26$ resonators and connected to \fiftyOhm~input-output ports. Dashed blue line: theoretical transmission of finite array without matching to \fiftyOhm~boundaries. Black line: theoretical transmission of finite array matched to \fiftyOhm~boundaries through two modified resonators at each boundary. Red line: measured transmission for a fabricated finite resonator array with boundary matching to input-output \fiftyOhm~coplanar waveguides. The measured ripple in transmission is less than $0.5$~dB in the middle of the passband. \textbf{e}, Measured group delay, $\taugroup$. Ripples in $\taugroup$ are less than $\delta\taugroup = 5$~ns in the middle of the passband.} 
\label{fig:LinearDesign}
\end{figure}

An optical and scanning electron microscope (SEM) image of the unit cell of the metamaterial slow-light waveguide used in this work are shown in Fig.~\ref{fig:LinearDesign}a.  The cell consists of a tightly meandered wire inductor section ($\meanind$; false color blue) and a top shunting capacitor ($\gndcap$; false color green), forming the lumped-element microwave resonator.  The resonator is surrounded by a large ground plane (gray) which shields the meander wire section. Laterally extended `wings' of the top shunting capacitor also provide coupling between the cells ($\coupcap$; false color green).  Note that at the top of the optical image, above each shunting capacitor, we have included a long superconducting island ($\qubitcap$; false color green); this is used in the next section as the shunting capacitance for Xmon qubits.  Similar lumped-element resonators have been realized with internal quality factors of $Q_i \sim 10^5$ and small resonator frequency disorder \cite{mirhosseini2018superconducting}, enabling propagation of light with low extinction from losses or disorder-induced scattering \cite{wiersma1997localization}. The waveguide resonators shown in Fig.~\ref{fig:LinearDesign}a have a bare resonance frequency of $\omegaMMres/2\pi\approx4.8$~GHz, unit cell length $d=290$~$\mu$m, and transverse unit cell width $w=540$~$\mu$m, achieving a compact planar form factor of $\dbar/\lambda = (\sqrt{dw})/(2\pi v/\omegaMMres)\approx 1/60$, where $v$ is the speed of light in a CPW on a infinitely thick silicon substrate. 

The unit cell is to a good approximation given by the electrical circuit shown in Fig.~\ref{fig:LinearDesign}b, in which the photon hopping rate is $J~\propto ~\coupcap/\gndcap$~\cite{girvin2011circuit}. We chose a ratio of $\coupcap/\gndcap \approx 1/70$, which yields a delay per resonator of roughly $2$~ns. Note that we have achieved this compact form factor and large delay per resonator while separating different lumped-element components by large amounts of ground plane, which minimizes spurious crosstalk between different unit cells. Analysis of the periodic circuit's Hamiltonian and dispersion can be found in App.~\ref{App:Theory}, where the dispersion is shown to be $\omega_k = \omegaMMres/\sqrt{1+4\frac{\coupcap}{\gndcap}\sin^2(kd/2)}$.  Figure~\ref{fig:LinearDesign}c shows a plot of the theoretical waveguide dispersion for an infinitely periodic waveguide, where the frequency of the bandedges of the passband are denoted with the circuit parameters of the unit cell. 

For finite resonator arrays care must be taken to avoid reflections at the boundaries that would result in spurious resonances (see Fig.~\ref{fig:LinearDesign}d, dashed blue curve, for example). To avoid these reflections, we taper the impedance of the waveguide by slowly shifting the capacitance of the resonators at the boundaries. In particular, we modify the first two unit cells at each boundary, but in principle, more resonators could have been modified for a more gradual taper. Increasing $\coupcap$ to increase the coupling between resonators, and decreasing $\gndcap$ to compensate for resonance frequency changes, effectively impedance matches the Bloch impedance of the periodic structure in the passband to the characteristic impedance of the input-output waveguides~\cite{pozar2009microwave}. In essence, this tapering achieves strong coupling of all normal modes of the finite structure to the input-output waveguides by adiabatically transforming guided resonator array modes into guided input-output waveguide modes. This loading of the normal modes lowers their $Q$ such that they spectrally overlap and become indistinguishable, changing the DOS of a finite array from that of a multi-mode resonator to that of finite-bandwidth continuum with singular bandedges. Further details of the design of the unit cell and boundary resonators can be found in App.~\ref{App:Taper}.  

\begin{figure*}[tbp]
\centering
\includegraphics[width = \textwidth]{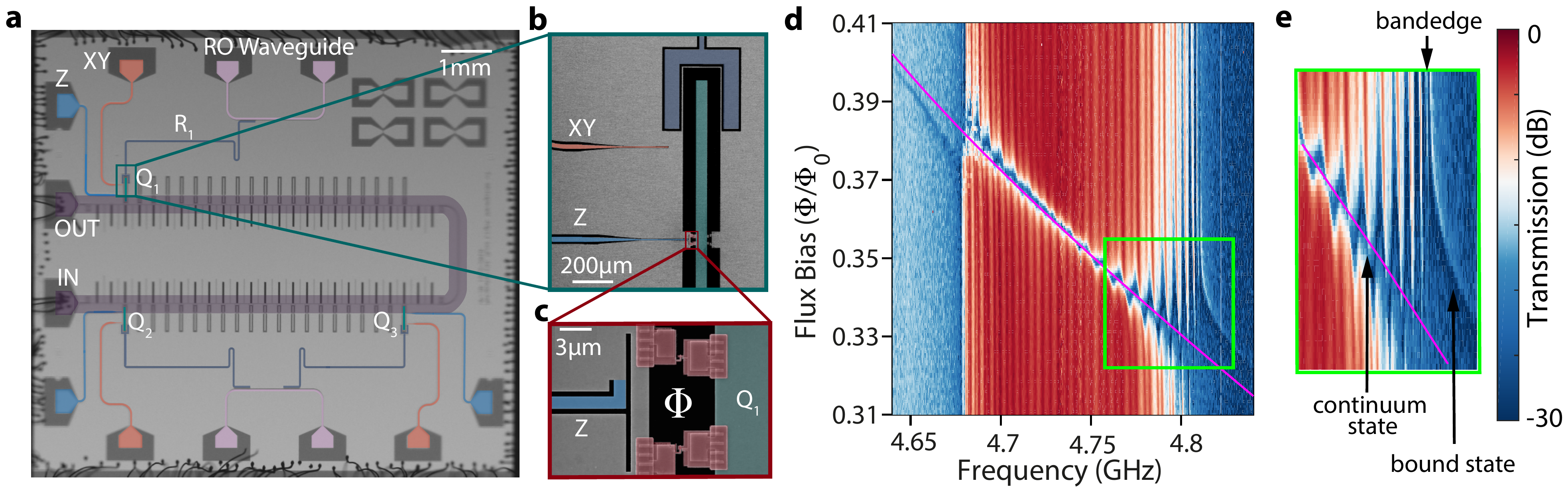}
\caption{\textbf{Artificial Atom Coupled to a Structured Photonic Reservoir.} \textbf{a}, False-colored optical image of a fabricated sample consisting of three transmon qubits (\Qone,$\text{Q}_2$,$\text{Q}_3$) coupled to a slow-light metamaterial waveguide composed of a coupled microwave resonator array. Each qubit is capacitively coupled to a readout resonator (false color dark blue) and a XY control-line (false color red), and inductively coupled to a Z flux-line for frequency tuning (false color light blue). The readout resonators are probed through feed-lines (false color lilac). The metamaterial waveguide path is highlighted in false color dark purple. \textbf{b}, SEM image of the \Qone~qubit, showing the long, thin shunt capacitor (false color green), XY control-line, the Z flux-line, and coupling capacitor to the readout resonator (false color dark blue). \textbf{c}, SEM zoom-in image of the Z flux-line and superconducting quantum interference device (SQUID) loop of \Qone~qubit, with Josephson Junctions and its pads false colored in crimson. \textbf{d}, Transmission through the metamaterial waveguide as a function of flux. The solid magenta line indicates the expected bare qubit frequency in the absence of coupling to the metamaterial waveguide, calculated based on the measured qubit minimum/maximum frequencies and the extracted anharmonicity. \textbf{e}, Zoom-in of transmission near the upper bandedge, showing the hybridization of the qubit with the bandedge, and its decomposition into a bound state in the upper bandgap and a radiative state in the continuum of the passband.} 
\label{fig:ArraywithQubits}
\end{figure*}

Using the above design principles, we fabricated a capacitively coupled $26$-resonator array metamaterial waveguide. The waveguide was fabricated using electron-beam deposited aluminum (Al) on a silicon substrate and was measured in a dilution refrigerator; transmission measurements are shown in Fig.~\ref{fig:LinearDesign}d,e, and further details of our fabrication methods and measurement set-up can be found in App.~\ref{App:Fab_Meas}. We find less than $0.5$~dB ripple in transmitted power and less than $10\%$ variation in the group delay ($\taugroup \equiv -\frac{d \phi}{d\omega}$, $\phi =\text{arg}(t(\omega))$, where $t$ is transmission) across $80$~MHz of bandwidth in the center of the passband, ensuring low distortion of propagating signals. Qualitatively, this small ripple demonstrates that we have realized a resonator array with small disorder and precise modification of the boundary resonators. More quantitatively, from the transmitted power measurements we extract a standard deviation in the resonance frequencies of $3\times 10^{-4} \times \omegaMMres$ (see App.~\ref{App:Disorder}). Furthermore, we achieve $\taudelay \approx 55$~ns of delay across the $1$~cm metamaterial waveguide, corresponding to a slow-down factor given by the group index of $n_g \approx 650$. We stress that this group delay is obtained across the center of the passband, rather than near the bandedges where large (and undesirable) higher-order dispersion occurs concomitantly with large delays.  

\section{Non-Markovian Radiative Dynamics}
\label{nonmarkovian}
In order to study the non-Markovian radiative dynamics of a quantum emitter, a second sample was fabricated with a metamaterial waveguide similar to that in the previous section, this time including three flux-tunable Xmon qubits~\cite{Barends2013} coupled at different points along the waveguide (see Fig.~\ref{fig:ArraywithQubits}a-c).  Each of the qubits is coupled to its own XY control line for excitation of the qubit, a Z control line for flux tuning of the qubit transition frequency, and a readout resonator (R) with separate readout waveguide (RO) for dispersive read-out of the qubit state. The qubits are designed to be in the transmon-limit~\cite{Koch2007} with large tunneling to charging energy ratio (see Refs.~\cite{keller2017transmon,mirhosseini2018superconducting} for further qubit design and fabrication details).  As in the test waveguide of Fig.~\ref{fig:LinearDesign}, the qubit-loaded metamaterial waveguide is impedance-matched to input-output \fiftyOhm~CPWs. In order to extend the waveguide delay further, however, this new waveguide is realized by concatenating two of the test metamaterial waveguides together using a CPW bend and internal impedance matching sections.  The Xmon qubit capacitors were designed to have capacitive coupling to a single unit cell of the metamaterial waveguide, yielding a qubit-unit cell coupling of $\gvacMMuc \approx 0.9J$. 

In this work only one of the qubits, \Qone, is used to probe the non-Markovian emission dynamics of the qubit-waveguide system.  The other two qubits are to be used in a separate experiment, and were detuned from \Qone~by approximately $ 1~\text{GHz}$ for all of the measurements that follow. At zero flux bias (i.e., maximum qubit frequency), the measured parameters of \Qone~are: $\omegage/2\pi=5.411 $~GHz, $\annharm/2\pi = (\omegaef -\omegage)/2\pi = -235$~MHz, $\omegares/2\pi = 5.871$~GHz, and $\gvacres/2\pi=88$~MHz.  Here, $\ket{g}$, $\ket{e}$, and $\ket{f}$ are the vacuum, first excited, and second excited states of the Xmon qubit, with $\omegage$ the fundamental qubit transition frequency, $\omegaef$ the first excited state transition frequency, and $\annharm$ the anharmonicity.  $\omegares$ is the readout resonator frequency, and $\gvacres$ is the bare coupling rate between the qubit state and the readout resonator. 

As an initial probe of qubit radiative dynamics, we spectroscopically probed the interaction of \Qone~with the structured 1D continuum of the metamaterial waveguide.  These measurements are performed by tuning $\omegage$ into the vicinity of the passband and measuring the waveguide transmission spectrum at low power.  A color intensity plot of the measured transmission spectrum versus flux bias used to tune the qubit frequency is displayed in Fig.~\ref{fig:ArraywithQubits}d.  These spectra show a clear anti-crossing as the qubit is tuned towards either bandedge of the passband (a zoom-in near the upper bandedge of the passband is shown in Fig.~\ref{fig:ArraywithQubits}e). As has been shown theoretically~\cite{john1994spontaneous,shen2019non}, in the single excitation manifold the interaction of the qubit with the waveguide results in a pair of qubit-photon dressed states of the hybridized system, with one state in the passband (a delocalized `continuum' state) and one state in the bandgap (a localized `bound' state).  This arises due to the large peak in the photonic DOS at the bandedge (in the lossless case, a van Hove singularity), the modes of which strongly couple to the qubit with a coherent interaction rate of $\OmegaWG \approx (\gvacMMuc^4/4J)^{1/3}$, resulting in a dressed-state splitting of $2\OmegaWG$.  This splitting has been experimentally shown to be a spectroscopic signature of a non-Markovian interaction between an emitter and a photonic crystal reservoir~\cite{sundaresan2019interacting,liu2017quantum}.  Further details and discussion can be found in App.~\ref{App:Theory}. 

The dressed state with frequency in the passband is a radiative state which is responsible for decay of the qubit into the continuum~\cite{john1991quantum}. On the other hand, the state with frequency in the gap is a qubit-photon bound state, where the qubit is self-dressed by virtual photons that are emitted and re-absorbed due to the lack of propagating modes in the waveguide for the radiation to escape. This bound state assumes an exponentially shaped photonic wavefunction of the form $\sum_x e^{-\abs{x}/\lambda}\hat{a}_x^\dagger \ket{\text{vac}}$, where $\ket{\text{vac}}$ is the state with no photons in the waveguide, $\hat{a}_x^\dagger$ is the creation operator of a photon in unit cell at position $x$ (with the qubit located at $x=0$), and $\lambda \approx \sqrt{J/(E_b-\omegaMMres)}$ is the state's localization length. In the theoretical limit of an infinite array, and in absence of intrinsic resonator and qubit losses, the qubit component of the bound state does not decay even though it is hybridized with the waveguide continuum; a behavior distinct from conventional open quantum systems. Practically, however, intrinsic losses and the overlap between the bound state's photonic wavefunction and the input-output waveguides will result in decay of the qubit-photon bound state. 


\begin{figure}[tbp]
\centering
\includegraphics[width = \columnwidth]{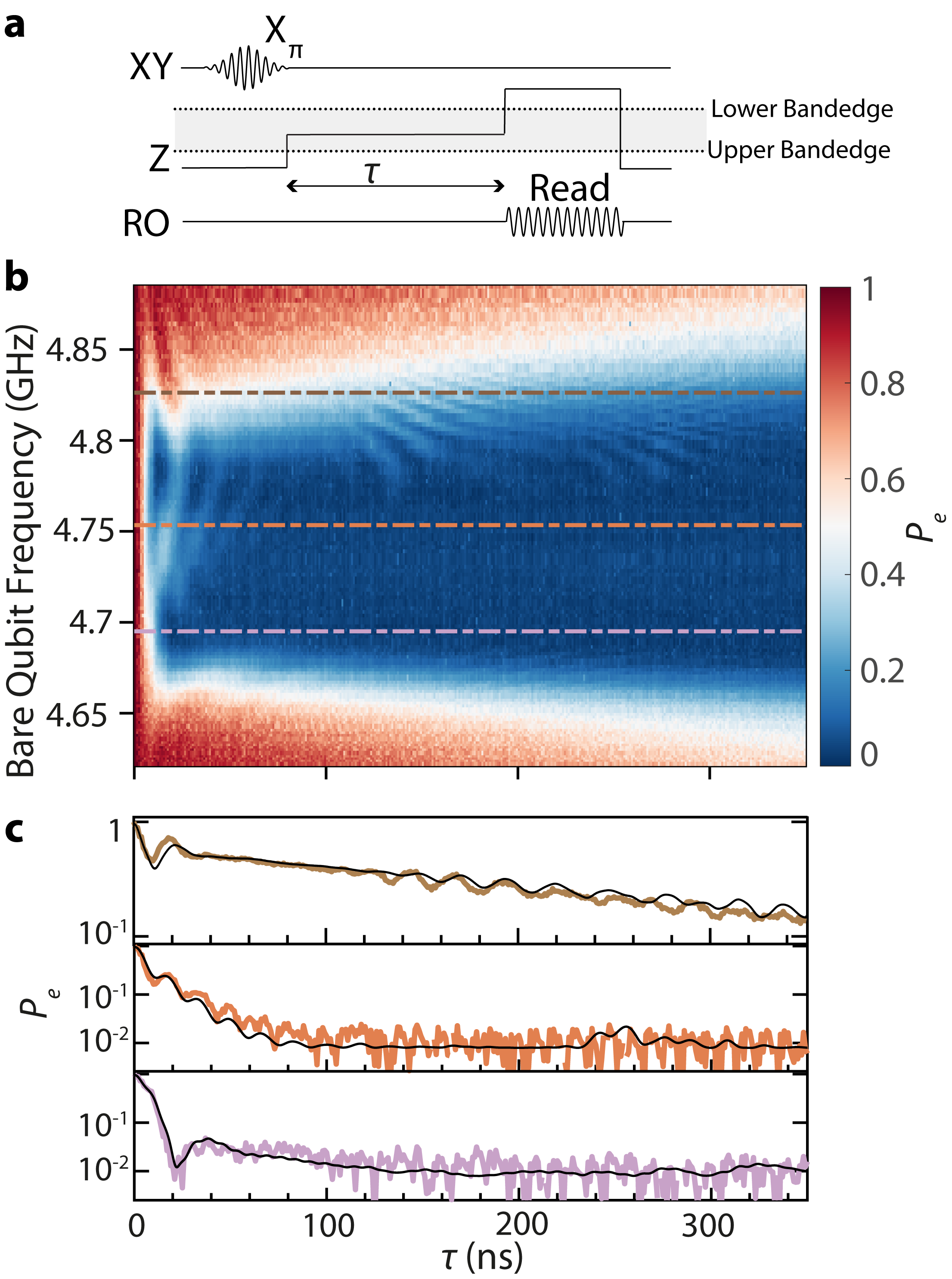}
\caption{\textbf{Non-Markovian Radiative Dynamics in a Structured Photonic Reservoir.} \textbf{a}, Pulse sequence for the time-resolved measurement protocol. The qubit is excited while its frequency is $250$~MHz above the upper bandedge, and then it is quickly tuned to the desired frequency ($\omegageint$) for a interaction time $\tau$ with the reservoir. After interaction, the qubit is quickly tuned below the lower bandedge for dispersive readout.  \textbf{b}, Intensity plot showing the excited state population of the qubit versus interaction time with the metamaterial waveguide reservoir as a function of the bare qubit frequency. \textbf{c}, Line cuts of the intensity plot shown in (\textbf{b}), where the color of the plotted curve matches the corresponding horizontal dot-dashed curve in the intensity plot. Solid black lines are numerical predictions of a circuit
model with experimentally fitted device parameters and an assumed $0.8\%$ thermal qubit population (see App.~\ref{App:QubitModeling} for further details).} 
\label{fig:BandedgeDecay}
\end{figure}

In complement to spectroscopic probing of the qubit-reservoir system, and in order to directly study the population dynamics of the qubit-photon dressed states, we also performed time-domain measurements as shown in Fig.~\ref{fig:BandedgeDecay}. In this protocol (illustrated in Fig.~\ref{fig:BandedgeDecay}a) we excite the qubit to state $\ket{e}$ with a resonant $\pi$-pulse on the XY control line, and then rapidly tune the qubit transition frequency using a fast current pulse on the Z control line to a frequency ($\omegageint$) within, or in the vicinity of, the slow-light waveguide passband.  After an interaction time $\tau$, the qubit is then rapidly tuned away from the passband, and the remaining qubit population in $\ket{e}$ is measured using a microwave probe pulse (RO) of the read-out resonator which is dispersively coupled to the qubit.  The excitation of the qubit is performed far from the passband, permitting initialization of the transmon qubit whilst it is negligibly hybridized with the guided modes of the waveguide. Dispersive readout of the qubit population is performed outside of the passband in order to minimize the loss of population during readout.  Note that, as illustrated in Fig.~\ref{fig:BandedgeDecay}a, the qubit is excited and measured at different frequencies on opposite sides of the passband; this is necessary to avoid Landau-Zener interference~\cite{shevchenko2010landau}. 

Results of measurements of the time-domain dynamics of the qubit population as a function of $\omegageint$ (the estimated bare qubit frequency during interaction with the waveguide) are shown as a color intensity plot in Fig.~\ref{fig:BandedgeDecay}b.  In this plot we observe a $400$-fold decrease in the $1/e$ excited state lifetime of the qubit as it is tuned from well outside the passband to the middle of the slow-light waveguide passband, reaching a lifetime as short as $7.5$~ns.  Beyond the large change in qubit lifetime within the passband, several other more subtle features can be seen in the qubit population dynamics near the bandedges and within the passband.  These more subtle features in the measured dynamics show non-exponential decay, with significant oscillations in the excited state population that is a hallmark of strong non-Markovianity in quantum systems coupled to amplitude damping channels~\cite{laine2010measure,breuer2012foundations}.  

The observed qubit emission dynamics in this non-Markovian limit are best understood in terms of the qubit-waveguide dressed states.  Fast (i.e., non-adiabatic) tuning of the qubit in state $\ket{e}$ into the proximity of the passband effectively initializes it into a superposition of the bound and continuum dressed states. The observed early-time interaction dynamics of the qubit with the waveguide then originate from interference of the dressed states, which leads to oscillatory behavior in the qubit population analogous to vacuum-Rabi oscillations~\cite{agarwal1985vacuum}. The frequency of these oscillations is thus set by the difference in energy between the dressed states.  The amplitude of the oscillations, on the otherhand, quickly decay away as the energy in the radiative continuum dressed state is lost into the waveguide.

All of these features can be seen in Fig.~\ref{fig:BandedgeDecay}c, which shows plots of the measured time-domain curves of the qubit excited state population for bare qubit interaction frequencies near the top, middle, and bottom of the passband.  Near the upper bandedge frequency we observe an initial oscillation period as expected due to dressed state interference.  Once the continuum dressed state has decayed away, a slower decay region free of oscillations can be observed (this is due to the much slower decay of the remaining qubit-photon bound state).  Finally, around $\tau\approx 115$~ns, there is an onset of further small amplitude oscillations in the qubit population.  These late-time oscillations can be attributed to interference of the remaining bound state at the site of the qubit with weak reflections occurring within the slow-light waveguide of the initially emitted continuum dressed state.  The $115$~ns timescale corresponds to the round trip time between the qubit and the CPW bend that connects the two slow-light waveguide sections.  

In the middle of the passband, we see an extended region of initial oscillation and rapid decay, albeit of smaller oscillation amplitude.  This is a result of the much smaller initial qubit-photon bound state population when tuned to the middle of the passband.  Near the bottom of the passband we see rapid decay and a single period of a much slower oscillation.  This is curious, as the dispersion near the upper and lower bandedge frequencies of the slow-light waveguide is nominally equivalent.  Further modelling has shown this is a result of weak non-local coupling of the Xmon qubit to a few of the nearest-neighbour unit cells of the waveguide. Referring to Fig.~\ref{fig:LinearDesign}c, the modes near the lower bandedge occur at the X-point of the Brillouin zone edge where the modes have alternating phases across each unit cell, thus extended coupling of the Xmon qubit causes cancellation-effects which reduces the qubit-waveguide coupling at the lower frequency bandedge.  Further detailed numerical model simulations of our qubit-waveguide system, and the correspondence between the observed dynamics and the theory of spontaneous emission by a two level system near a photonic bandedge~\cite{john1994spontaneous}, are given in App.~\ref{App:QubitModeling}.

\begin{figure}[tbp]
\centering
\includegraphics[width = \columnwidth]{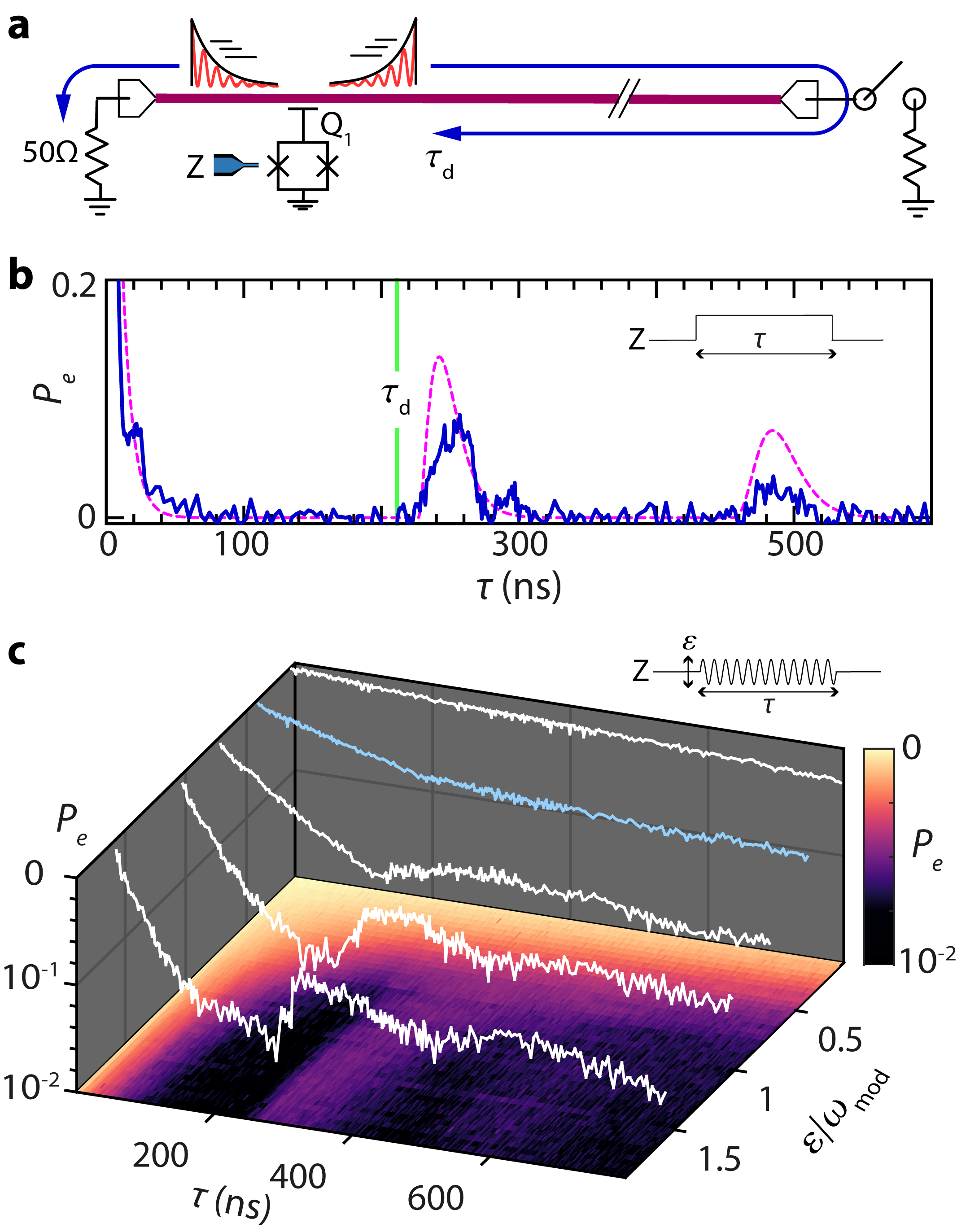}
\caption{\textbf{Time-Delayed Feedback from a Slow-Light Reservoir with a Reflective Boundary} \textbf{a}, Illustration of the experiment, showing the qubit coupled to the metamaterial waveguide which is terminated on one end with a reflective boundary via a microwave switch. \textbf{b}, Measured population dynamics of the excited state of the qubit when coupled to the metamaterial waveguide terminated in a reflective boundary.  Here the bare qubit is tuned into the middle of the passband. The onset of the population revival occurs at $\tau=227$~ns, consistent with round-trip group delay $(\taudelay)$ measurements at that frequency, while the emission lifetime of the qubit is $(\GammaoneD)^{-1} = 7.5$~ns. The magenta curve is a theoretical prediction for emission of a qubit into a dispersionless, lossless semi-infinite waveguide with equivalent $\taudelay$. \textbf{c}, Population dynamics under parametric flux modulation of the qubit, for varying modulation amplitudes, demonstrating a Markovian to non-Markovian transition. When the modulation index ($\epsilon/\omegamod$) is approximately 0.4 we have $\GammaoneD(\epsilon)=1/\taudelay$; the corresponding dynamical trace is colored in blue.}
\label{fig:Mirror}
\end{figure}

\section{Time-Delayed Feedback}
\label{time_delayed_feedback}
In order to further study the late-time, non-Markovian memory effects of the qubit-waveguide dynamics, we also perform measurements in which the end of the waveguide furthest from qubit Q$_{1}$ is terminated with an open circuit, effectively creating a `mirror' for photon pulses stored in the slow-light waveguide reservoir. As illustrated in Fig.~\ref{fig:Mirror}a, we achieve this \textit{in situ} by connecting the input microwave cables of the dilution refrigerator to the waveguide via a microwave switch.  The position of the switch, electrically closed or open, allows us to study a truly open environment for the qubit or one in which delayed-feedback is present, respectively. 

Performing time-domain measurements with the mirror in place and with the qubit frequency in the passband, we observe recurrences in the qubit population at one and two times the round-trip time of the slow-light waveguide that did not appear in the absence of the mirror (see Fig.~\ref{fig:Mirror}b). The separation of timescales between full population decay of the qubit and its time-delayed re-excitation demonstrates an exceptionally long memory of the reservoir due to its slow-light nature, and places this experiment in the deep non-Markovian regime~\cite{Tufarelli2014}. The small recurrence levels as they appear in Fig.~\ref{fig:Mirror}b are not due to inefficient mirror reflection, but rather can be explained as follows. 
Because the qubit emits towards both ends of the waveguide, half of the emission is lost to the unterminated end, while the other half is reflected by the mirror and returns to the qubit. In addition, the exponentially decaying temporal profile of the emission leads to inefficient re-absorption by the qubit and further limits the recurrence (see, for instance, Ref.~\cite{wang2011efficient, stobinska2009perfect} for details). These two effects can be observed in simulations of a qubit coupled to a dispersion-less and loss-less waveguide (pink dotted line). The remaining differences between the simulation and the measured population recurrence (blue solid line) can be explained by the effects of propagation loss and pulse distortion due to the slow-light waveguide's dispersion.

We also further probed the dependence of this phenomenon on the strength of coupling to the waveguide continuum by parametric flux modulation of the qubit transition frequency~\cite{li2013motional} when it is far detuned from the passband. This modulation creates sidebands of the qubit excited state, which are detuned from $\omega_{ge}$ by the frequency of the flux tone $\omegamod$. By choosing the modulation frequency such that a first-order sideband overlaps with the passband, the effective coupling rate of the qubit with the waveguide at the sideband frequency was reduced approximately by a factor of $\Besselone^2[\epsilon/\omegamod]$, where $\epsilon$ is the modulation amplitude and $\Besselone$ is a Bessel function of the first kind ($\epsilon/\omegamod$ is the modulation index). Keeping a fixed $\omegamod$, we observe purely exponential decay at small modulation amplitudes. However, above a modulation amplitude threshold we again observe recurrences in the qubit population at the round-trip time of the metamaterial waveguide, demonstrating a continuous transition from Markovian to non-Markovian dynamics. 

\section{Conclusion}
\label{conclusion}
In conclusion, we are able to observe non-Markovian dynamics of a Xmon qubit coupled to a 1D structured photonic reservoir realized by a metamaterial slow-light waveguide. In particular, near the bandedges, we observe non-exponential decay, which is due to the splitting of the qubit by the bandedge into a radiative state in the passband and a bound state outside of the passband. Moreover, by placing a reflective boundary on one end of the waveguide, we observe recurrences in the qubit population at the round-trip time of an emitted photon, as well as a Markovian to non-Markovian transition when varying the qubit-waveguide interaction strength. 

Our ability to achieve a true finite-bandwidth continuum with time-delayed feedback paves the way for several research avenues beyond the work presented here. In the short term, we envision probing our terminated waveguide-qubit system in the continuous, strongly-driven regime by tomography of the output radiation field, which will consist of a stream of strongly correlated photons with high entanglement dimensionality \cite{pichler2016photonic}. Furthermore, this output field has a direct mapping to continuous matrix product states, which can used for analog simulations of two-dimensional
interacting quantum fields, rather than one dimensional fields as has been previously done before~\cite{eichler2015exploring, barrett2013simulating}. And looking forward, additionally leveraging the multi-level structure of transmon-type qubits, by situating $\omega_{ef}$ in the passband and $\omegage$ in the gap, enables high fidelity generation of 2D cluster states for device parameters already achieved in this work\cite{pichler2017universal}. We therefore expect our results to find applications in future studies of non-Markovian open quantum systems,  studies of many-body physics, and measurement based quantum computation with microwave photons.

\begin{acknowledgments}
We thank Hannes Pichler for fruitful discussions regarding the mirror measurements, MIT Lincoln Laboratories for the provision of a traveling-wave parametric amplifier~\cite{Macklin2015} used for both spectroscopic and time-domain measurements in this work, Jen-Hao Yeh and Ben Palmer for the use of one of their cryogenic attenuators for reducing thermal noise in the metamaterial waveguide, and Hengjiang Ren and Xueyue Zhang for help during measurements, fabrication, and writing.  This work was supported by the AFOSR MURI Quantum Photonic Matter (grant 16RT0696), the Institute for Quantum Information and Matter, an NSF Physics Frontiers Center (grant PHY-1125565) with support of the Gordon and Betty Moore Foundation, and the Kavli Nanoscience Institute at Caltech. V.F gratefully acknowledges support from NSF GFRP Fellowship, and M.M (A.S.) gratefully acknowledges support from a KNI (IQIM) Postdoctoral Fellowship.
\end{acknowledgments}

\appendix

  


\section{Fabrication and Measurement Setup}
\label{App:Fab_Meas}

\subsection{Device Fabrication}

The devices used in this work were fabricated on 10 mm $\times$ 10 mm silicon substrates [Float zone (FZ) grown, 525 $\mu$m thickness, $>10 \text{k}\Omega$-cm resistivity], following similar techniques as in Ref.~\cite{keller2017transmon}. After standard solvent cleaning of the substrate, our first aluminum (Al) layer consisting of the ground plane, CPWs, metamaterial waveguide, and qubit capacitor was patterned by electron-beam lithography of our resist followed by electron-beam evaporation of $120$~nm aluminum at a rate of 1 nm/s. A liftoff process performed in n-methyl-2-pyrrolidone at $80$~$^{\circ}$C for $2.5$ hours (with 10 minutes of ultrasonication at the end) then yielded the aforementioned metal structures. 

In our qubit device, the Josephson junctions were fabricated using double-angle electron beam evaporation of $60$~nm and $120$~nm of Al (at 1 nm/s) on suspended Dolan bridges, with an intervening 20 minute oxidation and a subsequent 2 minute oxidation at 10 mbar, followed by liftoff as described above. Note that prior to the double-angle evaporation, the sample was cleaned by an oxygen plasma treatment and a HF vapor etch. Finally, in order to electrically connect the evaporated Josephson junctions to the first Al layer, a 6min argon ion mill was performed to locally remove surface aluminum oxide around the areas of overlap between the first Al layer and the Josephson junctions, which was followed by evaporation of an additional ``bandage” layer of $140$~nm Al that electrically connected the metal layers. Asymmetric Josephson junctions were fabricated in all qubits’ SQUID loops to reduce dephasing from flux noise, with a design ratio of the larger junction area to the smaller junction area of approximately 6. 

\subsection{Measurement Setup}

\begin{figure}[tbp]
\centering
\includegraphics[width = \columnwidth]{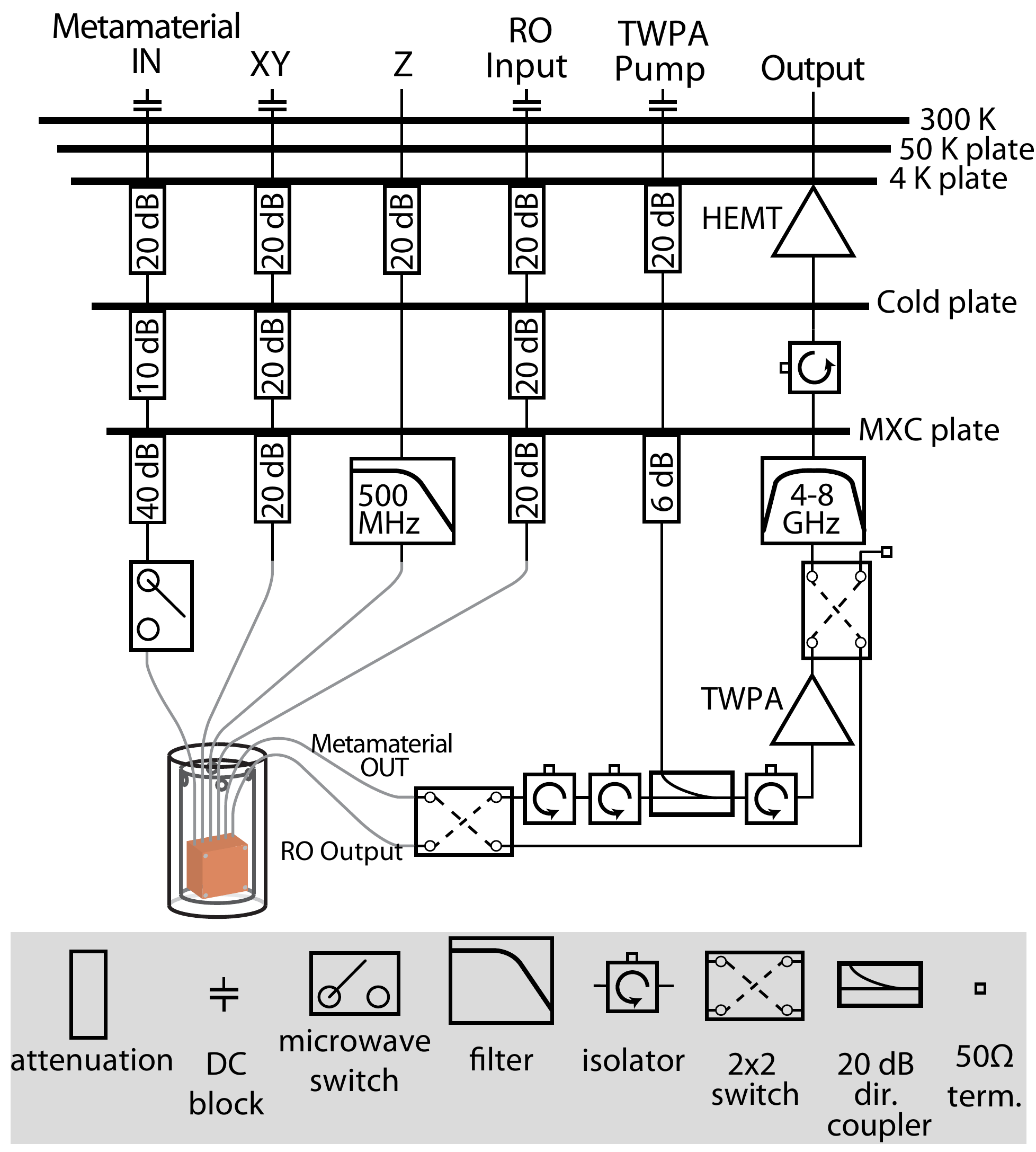}
\caption{Schematic of the measurement chain inside the dilution refrigerator. See Appendix text for further details (``dir." is shorthand for ``directional", and ``term." is shorthand for ``termination"). See Fig.~\ref{fig:ArraywithQubits} for electrical connections at the sample. } 
\label{fig:meas_setup}
\end{figure}

A schematic of the measurement chain used in this work is shown in Fig.~\ref{fig:meas_setup}.  Measurements were performed in a 3He/4He dry dilution refrigerator, with a base fridge temperature at the mixing chamber (MXC) plate of $\Tf = 12$~mK.  The waveguide sample was wire bonded to a CPW printed circuit board (PCB) with coaxial connectors, and housed inside a small copper box that is mounted to the MXC plate of the fridge.  The copper box and sample were mounted inside a cryogenic magnetic shield to reduce the effects of stray magnetic field. 

Attenuators were placed at several temperature stages of the fridge to provide thermalization of the coaxial input lines, and to reduce thermal microwave noise at the input to the sample. We used different attenuation configurations for our GHz microwave lines (Metamaterial IN, XY, RO Input, TWPA pump) as compared to our flux line (Z), with significantly less attenuation for the latter, for reasons explained in Ref. \cite{yeh2017microwave}. In addition, we included in the flux line a (reflective) low-pass filter, with corner frequency at $500$~MHz, to minimize thermal noise photons at higher frequencies while maintaining short rise and fall time of pulses for fast flux control. Also note that the $40$~dB attenuation of the ``Metamaterial IN" line at the MXC plate includes a $20$~dB thin-film ``cold attenuator" \cite{krinner2019engineering} to ensure a more complete reduction of thermal photons in the metamaterial waveguide. 

Our amplifier chain at the ``Output" line consisted of a travelling-wave parametric amplifier (TWPA) as the initial amplification stage \cite{Macklin2015}, followed by a CITCRYO4-12A high mobility electron transistor (HEMT) amplifier mounted at the 4K plate, and additional amplifiers at room temperature (Miteq AFS3-00101200-42-LN-HS, AMT A0262). For operation of the TWPA, a microwave pump signal was added to the amplifier via the coupled port of a 20dB directional coupler, with its isolated port terminated in \fiftyOhm. In between the two amplifiers, we have included a reflective bandpass filter (thermalized to the MXC plate) to suppress noise outside of $4$--$8$~GHz, and used superconducting NbTi cables to minimize loss from the MXC plate to the 4K plate. We have also included two isolators in between the directional coupler and the sample in order to shield the sample from the strong TWPA pump, as well as an isolator in between the TWPA and the directional coupler in order to suppress any standing waves between the two elements due to spurious impedance mismatches; our isolators consist of 3 port circulators with the third port terminated in \fiftyOhm. All \fiftyOhm~terminations are rated for cryogenic operaion and are thermalized to the MXC plate in order to suppress thermal noise from their resistive elements.

We also employed microwave switches in our measurement chain in order to provide \textit{in situ} experimental flexibility in the following manner. As discussed in the main text, in between the ``Metamaterial IN" chain and the metamaterial waveguide we have placed a Radiall R573423600 microwave switch. By electrically opening the switch, we can establish an open circuit at the end of the waveguide furthest from \Qone, effectively creating a mirror for emission from \Qone~and thereby inducing time-delayed feedback. 

In addition, in order to utilize our amplifier chain for either spectroscopic or time-domain measurements within the same cool-down, we employed Radiall R577432000 2x2 microwave switches for selective routing of the outputs of the metamaterial waveguide or the readout waveguide to the amplification chain. With our switch configuration, we ensured that when routing the readout waveguide output to the amplification chain, the metamaterial waveguide output was connected to a \fiftyOhm~termination. This allowed us to  maintain a \fiftyOhm~environment at the metamaterial output at all times, and thereby ensured that the metamaterial waveguide remained an open quantum system due to its coupling to the \fiftyOhm~continuum of modes. By employing two 2x2 switches instead of one, we had the ability to bypass the TWPA amplifier if desired, although ultimately the TWPA was used when collecting all measurement data presented in Figs. \ref{fig:ArraywithQubits}--\ref{fig:Mirror}. 

 For spectroscopic measurements, the ``Metamaterial IN” and ``Output” lines were connected to the input and output of a ZNB20 Rohde \& Schwarz vector network analyzer (VNA), respectively. For time-domain measurements, GHz excitation and readout pulses were generated by upconversion of MHz IF in-phase (I) and quadrature (Q) signals sourced from a Keysight M320XA arbitrary waveform generator (AWG), utilizing a Marki IQ-4509 IQ mixer and a LO tone supplied by a BNC 845 microwave source. Following amplification, output readout signals were downconverted (using an equivalent mixer and the same LO source) and subsequently digitized using an Alazar ATS9371 digitizer. For all measurements, qubit flux biasing was also sourced from a M320XA AWG, the TWPA pump tone was sourced by an Agilent E8257D microwave source, and all inputs to the dilution refrigerator were low-pass filtered and attenuated such that the noise levels from the electronic sources were reduced to a $300$~K Johnson-Nyquist noise level.   
 
\section{Capacitively Coupled Resonator Array Waveguide Fundamentals}
\label{App:Theory}

\subsection{Band Structure Analysis}

We consider a periodic array of capacitively coupled LC resonators, with unit cell circuit diagram shown in the inset to Fig.~\ref{fig:LinearDesign}a. The Lagrangian for this system can be constructed as a function of node fluxes $\phi_x$ of the resonators, and is written as,

\begin{equation}
		L = \sum_{x}\left[\frac{1}{2}\gndcap\dot{\phi}_x^2 + \frac{1}{2}\coupcap(\dot{\phi}_x-\dot{\phi}_{x-1})^2 - \frac{{\phi_x}^2}{2L_0}\right].
\end{equation}

\noindent Since we seek traveling wave solutions to the problem, it is convenient to work with the Fourier transform of the node fluxes, defined as 

\begin{equation}
	\phi_k = \frac{1}{\sqrt[]{M}}\sum_{x=-N}^{N} \phi_x e^{-ikxd},
\end{equation}

\noindent where $M=2N+1$ is the total number of periods of a structure with periodic boundary conditions, $d$ is the lattice constant of the resonator array, and $k$ are the discrete momenta of the first Brillouin zone's guided modes and are given by $k = \frac{2\pi m}{Md}$ for integer $m=[-N,N]$. Using the inverse Fourier transform,

\begin{equation}
	\phi_x = \frac{1}{\sqrt[]{M}}\sum_{k} \phi_k e^{ikxd},
\end{equation}

\noindent we arrive at the following $k$-space Lagrangian

\begin{equation}
	L = \sum_{k}\left[\frac{1}{2}\gndcap\dot{\phi}_k \dot{\phi}_{-k} + \frac{1}{2}\coupcap\dot{\phi}_k\dot{\phi}_{-k}\left| 1-e^{-ikd}\right|^2  - \frac{\phi_k \phi_{k}}{2L_0}\right], 
\end{equation}

\noindent where we note that $\left| 1-e^{-ikd}\right|^2$ is equivalent to $4\sin^2{\left(kd/2\right)}$. We then obtain the Hamiltonian via the standard Legendre transformation using the canonical node charges $	Q_k = \frac{\partial{L}}{\partial{\dot{\phi}_k}}= \dot{\phi}_{-k}\left(\gndcap + 4\coupcap\sin^2{\left(kd/2\right)}\right)$, yielding:

\begin{equation} \label{H_flux_charge}
	H = \sum_{k}\left[\frac{1}{2} \frac{Q_k Q_{-k}}{\left(4\coupcap\sin^2{\left(kd/2\right)}+\gndcap\right)} + \frac{\phi_k \phi_{-k}}{2L_0}\right].
\end{equation}

Promoting charge and flux to quantum operators and utilizing the canonical commutation relation $\left[\hat{\phi_k},\hat{Q_{k'}} \right] = i\hbar \delta_{kk'}$, we define the following creation and annihilation operators:

\begin{equation}
\begin{split}
\loweringk &= \sqrt{\frac{m_k \omega_k}{2\hbar}} \left( \hat{\phi}_k +\frac{i}{m_k \omega_k} \hat{Q}_{-k}  \right), \\
\raisingk &= \sqrt{\frac{m_k \omega_k}{2\hbar}} \left( \hat{\phi}_{-k} +\frac{i}{m_k \omega_k} \hat{Q}_{k}  \right),
\end{split}
\end{equation}

\noindent where $m_k = \left(\gndcap + 4\coupcap\sin^2{\left(kd/2\right)} \right)$.  The resulting dispersion relation, $\omega_k$, plotted in Fig.~\ref{fig:LinearDesign}c is given by,

\begin{equation}\label{dispersion}
    \omega_k = \frac{\omegaMMres}{\sqrt{1+4\frac{\coupcap}{\gndcap}\sin^2(kd/2)}}, 
\end{equation} 

\noindent where $\omegaMMres=1/\sqrt{L_0C_0}$, and $\left[\loweringk,\hat{a}^\dagger_{k'} \right] = \delta_{kk'}$. Expressing the flux and charge operators in terms of $\loweringk,\hat{a}^\dagger_{k'}$ and substituting them into Eq.~(\ref{H_flux_charge}), we recover the second-quantized Hamiltonian in the diagonal k-space basis

\begin{equation} \label{diag_H}
    \hat{H} = \sum_{k }\hbar  \omega_k \left(\frac{1}{2}+ \raisingk\loweringk \right).
\end{equation}

\noindent Note that, given the translational invariance of the capacitively coupled resonator array circuit, it was expected that the the Hamiltonian would be diagonal in the Fourier plane-wave basis (Bloch Theorem). 

Also note that, for two capacitively coupled LC resonators, their coupling $J=\frac{\omegaMMres}{2}(\coupcap/(\gndcap+\coupcap))$ is positive-valued~\cite{girvin2011circuit} due to the fact that the anti-symmetric odd mode of the circuit is the lower energy eigenmode. This results in positive-valued photon hopping terms in the Hamiltonian, which directly lead to a maximum in frequency at the  $\Gamma$ point and opposite directions of the phase velocity and group velocity in the structure, as observed in other dispersive media~\cite{gersen2005direct,wang2006direct,woodley2006backward}.

\subsection{Comparison to Tight-Binding Model}

In the limit $\gndcap \gg \coupcap$ the dispersion is well approximated to first order by a tight-binding model with dispersion given by $\omega_k= \omega_p + 2J\cos \left (kd \right)$, where $J=\omegaMMres(\coupcap/2\gndcap)$ is approximately the nearest-neighbour coupling between two resonators of the resonator array, and $\omega_p=(\omegaMMres-2J)$ is the center of the passband. The difference in the two dispersion relations reflects the coupling beyond nearest-neighbor that arises due to the topology of the circuit, in which any two pairs of resonators are electrically connected through some capacitance network dependent on their distance. The magnitude of these interactions is captured in the Fourier transform of the dispersion. Consider the Fourier transform for the annihilation operators of the (localized) mode of the individual resonator located at position $x$,

\begin{equation} \label{fourier_a}
	\loweringk = \frac{1}{\sqrt[]{M}}\sum_{x} \hat{a}_x e^{-ikxd}.
\end{equation}

\noindent Substituting Eq.~(\ref{fourier_a}) into Eq.~(\ref{diag_H}), we arrive at the following real-space Hamiltonian,

\begin{equation} \label{realSpace_H}
    \hat{H} = \hbar\sum_{x}\sum_{x'} V(x-x')\hat{a}_x^\dagger\hat{a}_{x'}, 
\end{equation}

\noindent where $V(x-x')$ is the distance-dependent interaction strength between two resonators located at positions $x$ and $x'$, and is simply given by the Fourier transform of the dispersion relation,

\begin{equation} \label{Fourier_disp}
	V(x-x') = \frac{1}{M}\sum_{k} \omega_k e^{-ikd(x-x')}.
\end{equation}

\noindent For example, substituting the tight-binding dispersion $\omega_k=\omega_p+2J\cos \left (kd \right)$ into Eq.~(\ref{Fourier_disp}) yields $V(x-x')= \omega_p \delta_{x,x'} + 2J\left(\delta_{x-x',1} + \delta_{x-x',-1}\right)$, which, upon substitution into Eq.~(\ref{realSpace_H}), recovers the tight-binding Hamiltonian with only nearest-neighbor coupling. 

\begin{figure}[tbp]
\centering
\includegraphics[width = \columnwidth]{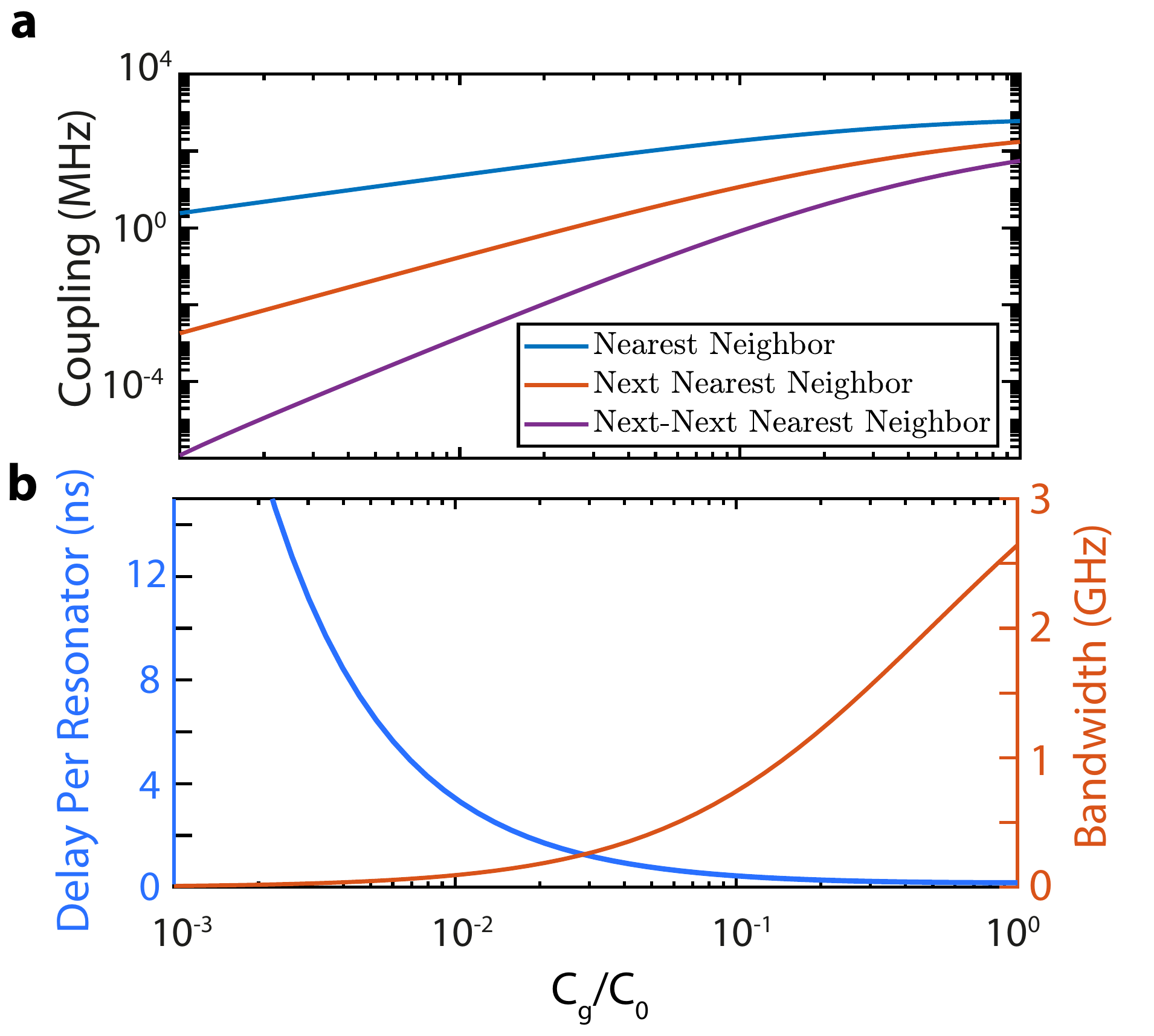}
\caption{\textbf{a}. Magnitude of nearest neighbor, next-nearest neighbor, and next-next nearest neighbor inter-resonator couplings in an (infinite) capacitively coupled resonator array as a function of $\coupcap/\gndcap$ ratio. The bare resonator frequency was chosen to be 4.8GHz. \textbf{b} Magnitude of delay per resonator and bandwidth of the passband as a function of $\coupcap/\gndcap$ ratio. The bare resonator frequency was again chosen to be 4.8GHz}
\label{fig:couplings_delay}
\end{figure}

In Fig.~\ref{fig:couplings_delay}a we plot the magnitudes of nearest neighbor ($x-x'=1$), next-nearest neighbor ($x-x'=2$), and next-next-nearest neighbor ($x-x'=3$) couplings in the capacitively coupled resonator array as a function of $\coupcap/\gndcap$, calculated numerically via the discrete Fourier transform of the dispersion relation. It is evident that for small $\coupcap/\gndcap$ the nearest neighbor coupling overwhelmingly dominates. 

\subsection{Qubit Coupled to Passband of a Waveguide}

The Hamiltonian of a transmon-like qubit coupled to the metamaterial waveguide via a single unit cell, where only the first two levels of the transmon ($\ket{g}, \ket{e}$) are considered, can be written as ($\hbar = 1, d=1$),

\begin{equation}
	H = \omega_{ge}\ket{e}\bra{e} + \sum_{k }\omega_k \raisingk\loweringk + \frac{\gvacMMuc}{\sqrt[]{M}}\sum_k \left (\raisingk \loweringq + \loweringk \raisingq \right),
\label{qubit_H_SI}
\end{equation}

\noindent where $\omega_k$ is given by Eq.~(\ref{dispersion}). For an infinite array, the time-independent Schrodinger equation $\hat{H}\ket{\psi} = E\ket{\psi}$ has two types of solutions in the single photon manifold: there are scattering eigenstates, which have an energy within the passband, and there are bound states that are energetically separated from the passband continuum. We demonstrate this in the following analysis. First, we substitute into $\hat{H}\ket{\psi} = E\ket{\psi}$ the following ansatz for the quantum states of the composite qubit-waveguide system, i.e. for dressed states of the qubit,

\begin{equation}
	\ket{\psi} = c_e\ket{e,\text{vac}} + \sum_k  c_k \raisingk \ket{g,\text{vac}},
\label{ansatz}
\end{equation}

\noindent where $\ket{\text{vac}}$ corresponds to no excitations in the waveguide. Doing this substitution and subsequently collecting terms, we arrive at the following coupled equations for $c_e$ and $c_k$:

\begin{align}
\label{coupled1}
c_e & = \frac{\gvacMMuc}{\sqrt[]{M}}\sum_k  \frac{c_k}{E-\omegage},  \\
c_k & =  \frac{\gvacMMuc}{\sqrt[]{M}} \frac{c_e}{E-\omega_k}. \label{coupled2}
\end{align}

\noindent By further assuming that the waveguide supports a continuum of modes (which is appropriate for a finite tapered waveguide, as described in the main text), the sum can be changed into an integral $\sum_k \rightarrow \frac{1}{\Delta_k} \sum_k \Delta_k \rightarrow \frac{1}{\Delta_k} \int_{-\pi}^\pi \dk$, where $\Delta_k = 2\pi/M$. In this continuum limit, $E$ can be found by first substituting Eq.~(\ref{coupled2}) into Eq.~(\ref{coupled1}) and subsequently dividing both sides by $c_e$, which yields the following transcendental equation for $E$,

\begin{equation}
	E = \omega_{ge} + \frac{1}{2\pi} \int \dk \frac{\gvacMMuc^2}{E-\omega_k},
\label{transcandental_SI}
\end{equation}

\noindent where the integral on the right-hand side of Eq.~(\ref{transcandental_SI}) is known as the ``self-energy" of the qubit 
\cite{john1994spontaneous,gonzalez2017markovian,lambropoulos2000fundamental}. Note that in the opposite limit of a single resonator (where $\omega_k$ takes on a single value and the density of states $\frac{\partial{\omega}}{\partial{k}}$ becomes a delta--function at that value), Eq.~(\ref{transcandental_SI}) yields the familiar Jaynes-Cummings  splitting $\sqrt{\delta^2 + \gvacMMuc^2}$. 

Computation of the self-energy for $E$ such that $E>\omega_k~\text{or}~E<\omega_k~~\forall k$, i.e. for energies outside of the passband, yields real solutions for Eq.~(\ref{transcandental_SI}). On the other hand, for energies $E$ inside the passband, the self-energy integral contains a divergence at $E=\omega_k$ for real $E$ while there is no divergence if $E$ is allowed to be complex with an imaginary component; thus Eq.~(\ref{transcandental_SI}) has complex solutions when $\text{Re}(E)$ is inside the passband. While a Hermitian Hamiltonian such as the one in Eq.~(\ref{qubit_H_SI}) by definition does not contain complex eigenvalues, it can be shown that the magnitude of the imaginary component of complex solutions of Eq.~(\ref{transcandental_SI}) gives the decay rate of an excited qubit for a qubit dressed state with energy in the passband.  For further details we suggest Refs.~\cite{gonzalez2017markovian,lambropoulos2000fundamental,john1991quantum} to the reader. Thus, the existence of complex solutions of Eq.~(\ref{transcandental_SI}) reflect the fact that qubit dressed states with energy in the passband are radiative states that decay into the continuum, characteristic of open quantum systems coupled to a continuum of modes. In contrast, the dressed states with (real) energies outside of the passband do not decay, and are known as qubit-photon bound states in which the photonic component of the dressed state wavefunction remains bound to the qubit and is not lost into the continuum. 

For further analytical progress, we consider only the upper bandedge, and make the effective-mass approximation. This approximation is tantamount to assuming the dispersion is quadratic, such that $\omega_k \approx \omegaMMres - Jk^2$, which is obtained in the limit of small $C_g/C_0$ (where $\omega_k$ is well approximated by the tight binding cosine dispersion) and small $k$ (where $\cos(k)$ to second order is approximately $1 - k^2/2$). This approximation is appropriate when $\omegage$ is close to the upper bandedge, where the qubit is dominantly coupled to the $\Gamma$-point $k=0$ modes close to the bandedge due to the van Hove singularity in the DOS, and when the lower bandedge is sufficiently detuned from the qubit. Complimentary analysis for the lower bandedge can also be done in the same manner. For a more detailed derivation, see Refs.~\cite{calajo2016atom,lombardo2014photon,gonzalez2017markovian}.

Under the effective-mass approximation, the self-energy integral in Eq.~(\ref{transcandental_SI}) can be easily analyzed by taking the bounds of integration to infinity, and is calculated to be $\gvacMMuc^2/2\sqrt{J(E-\omegaMMres)}$. For $\omegage = \omegaMMres$,  Eq.~(\ref{transcandental_SI}) then has the following two solutions:

\begin{align}
\label{E_b}
E_b &= \omegaMMres + (\gvacMMuc^4/4J)^{1/3},  \\
E_r &= \omegaMMres- e^{i\pi/3}(\gvacMMuc^4/4J)^{1/3} \label{E_r}.
\end{align}

\noindent These two solutions are indicative of a splitting of the qubit transition frequency by the bandedge into two dressed states: a radiative state with energy $E_r$ in the passband and a bound state with energy $E_b$ above the bandedge. The magnitude difference between the dressed state energies is $2(\gvacMMuc^4/4J)^{1/3}$, which is the frequency of coherent qubit-to-photon oscillations for an excited qubit at the photonic bandedge. 

For the remainder of the analysis, we focus on the qubit-photon bound state of the system. The wavefunction of the bound state with energy $E$ can be obtained by first substituting Eq.~(\ref{coupled2}) into Eq.~(\ref{ansatz}), which yields

\begin{equation}
\ket{\psi_E} = c_e \left (\ket{e} + \frac{\gvacMMuc}{\sqrt[]{M}}\sum_k \frac{ 1}{E-\omega_k} \raisingk \ket{g,\text{vac}} \right ).
\label{eigenstates}
\end{equation}

\noindent The qubit and photonic components of the bound state can be calculated from the normalization condition of $\ket{\psi_E}$, 

\begin{equation}
    \abs{c_e}^2 \left ( 1+ \frac{1}{2\pi} \int \dk \abs{\frac{ \gvacMMuc}{E-\omega_k}}^2 \right ) = 1. 
\label{c_e}
\end{equation}

\noindent By assuming $E>\omegaMMres$, the integral in Eq.~(\ref{c_e}) is calculated to be equal to $\gvacMMuc^2/4\sqrt{J(E-\omegaMMres)^3}$, which yields the following magnitude for the qubit component of the bound state,   

\begin{equation}
\abs{c_e}^2 = \left (1 + \frac{1}{2}\frac{E-\omegage}{E-\omegaMMres} \right)^{-1}, 
\label{abs_ce}
\end{equation}

\noindent whereas the photonic component is simply $\int \dk \abs{c_k}^2 = 1-\abs{c_e}^2$. We can thus see that when $E \approx \omegage \neq \omegaMMres$, the qubit is negligibly hybridized with the passband modes and $\abs{c_e}^2 \approx 1$. On the otherhand, as $\omegage \rightarrow \omegaMMres$ we have $\abs{c_e}^2 \rightarrow 2/3$, indicating that the bound-state photonic component contains half as much energy as the qubit component when the qubit is tuned to the bandedge. 

We can also obtain the real-space shape of the photonic bound state by inserting Eq.~(\ref{fourier_a}) into Eq.~(\ref{eigenstates}), where for a continuum of modes in $k$-space we arrive at the following photonic wavefunction, 

\begin{equation}
    \sum_x e^{-\abs{x}/\lambda}\hat{a}_x^\dagger \ket{g,\text{vac}},
\end{equation}

\noindent up to a normalization constant, where $\lambda = \sqrt{J/(E-\omegaMMres)}$ and the qubit is assumed to reside at $x=0$. We thus find an exponentially localized photonic wavefunction for the bound state. The localization length $\lambda$ increases as $J$ increases, indicating that the bound state becomes more delocalized across multiple resonators as the strength of coupling between the resonators in the waveguide increases, whereas $\lambda$ diverges as the $E \rightarrow \omega_0$, which is associated with full delocalization of the bound-state as its energy approaches the continuum of the passband. 

\subsection{Group Delay}
Lowering the ratio $\coupcap/\gndcap$ effectively lowers the photon hopping rate $J$ between resonators, and can thus be chosen to significantly decrease the group velocity of propagating modes of the structure, albeit at the cost of decreased bandwidth of the passband modes. The group velocity may be obtained from $\frac{\partial \omega_k}{\partial k}$, while the bandwidth can be calculated to be equal to $\omegaMMres \left(1-1/\sqrt{1+4\coupcap/\gndcap} \right)$; both are plotted in Fig.~\ref{fig:couplings_delay}b. Note that although the group velocity approaches zero near the bandedge, a traveling pulse at the bandedge frequency would experience significant distortion due to the rapidly changing magnitude of the group velocity near the bandedge.  At the center of the passband where the dispersion is nearly linear, however, it is possible to have propagation with minimal distortion. 

Hence, in order to effectively use the coupled resonator array as a delay line, the coupling should be made sufficiently high such that the bandwidth of propagating modes (where the dispersion is also nearly linear) is sufficiently high, and the effect of resonator frequency disorder due to fabrication imperfections is tolerable. After the resonator coupling constraints have been met, the desired delay may be achieved with a suitable number of resonators. It is thus evident that the ability to fabricate resonators of sub-wavelength size with minimal frequency disorder is critical to the effectiveness of implementing a slow-light waveguide with a coupled resonator array.

An appropriate metric to compare the performance of the resonator array as a delay line against dispersionless waveguides is to consider the delay achieved per area rather than per length, in order to account for the transverse dimensions of the resonators. In addition, typical implementations of delay lines with CPW geometries commonly require a high degree of meandering in order to fit in a packaged device; thus the pitch and turn radius of the CPW meandered trace also must be taken into account when assessing delay achieved per area. However, by making certain simplifying assumptions about the resonators it is possible to gain intuition on how efficient the resonator array is in achieving long delays compared to a dispersionless CPW. For the resonators implemented in the Main text (see Fig.~\ref{fig:LinearDesign}), the capacitive elements of the resonator are electrically connected to one end of the meander while the opposite end of the meander is shunted to ground. This geometry is therefore topologically similar to a $\lambda/4$ resonator, and consequently the lengths of the meander and a conventional $\lambda/4$ CPW resonator will be similar to within an order of magnitude for conventional implementations (here $\lambda$ is the wavelength of the CPW resonator mode). 

Thus, by approximating that a single resonator of the array occupies the same area as a $\lambda/4$-section of CPW, a direct comparison between the delays of the two different waveguides can be made. In the tight-binding limit, the group delay in the middle of the passband is approximately equal to $2J$, where $J$ is the coupling between two resonators of the array. Hence, for N resonators  $\taudelay^{\text{array}}/\taudelay^{\text{CPW}}=\frac{N/2J}{N\lambda/4c} \sim \omegaMMres/J$, where $\tau_d$ is group delay and $c$ is the group velocity of light in the CPW. Hence, the resonator array is more efficient as a delay line when compared to conventional CPW by a factor of approximately $\omegaMMres/J$ (assuming group velocity is approximately equal to phase velocity in the CPW).  In practice, this factor will also depend on the particular geometrical implementations of both kinds of waveguide. For example, for the resonator array described in Fig.~\ref{fig:LinearDesign}, $\omegaMMres/J \approx 120$ and $\taudelay = 55$~ns delay was achieved in the middle of the passband for a resonator array of area $A = 6$~mm$^2$.  This constitutes a factor of 60 (500) improvement in delay per area achieved over the CPW delay line in Ref.~\cite{zhong2019violating} (Ref.~\cite{wang2005wide}).

\section{Physical Implementation of Finite Resonator Array}
\label{App:Taper}

\begin{figure*}[tbp]
\centering
\includegraphics[width = \textwidth]{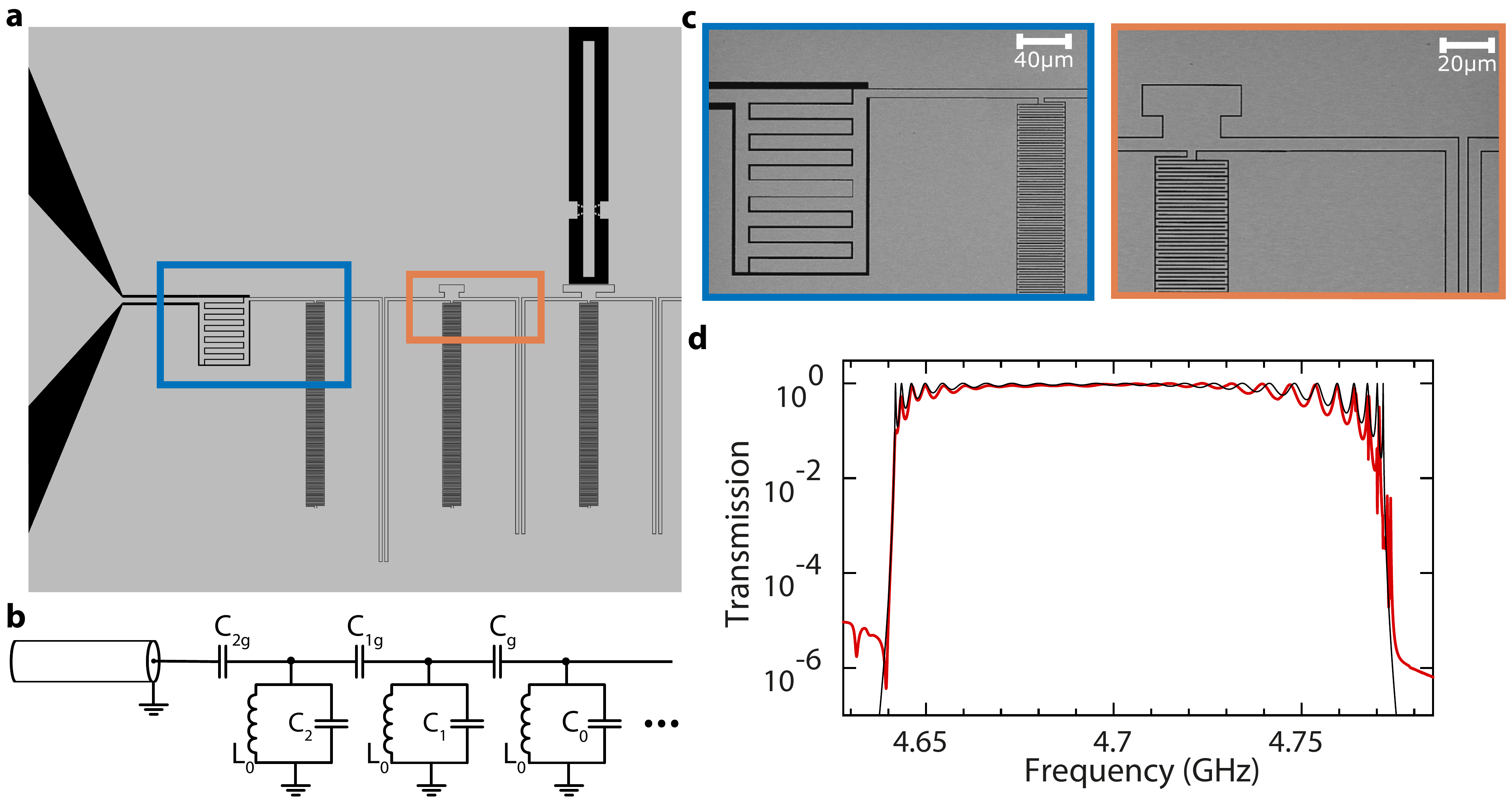}
\caption{\textbf{a}, CAD diagram showing the end of the finite resonator array, including boundary matching circuit (which in this case includes the first two resonators) and the first unit cell. \textbf{b}, Corresponding circuit model of the end of the finite resonator array. \textbf{c}, Zoomed-in SEM images of the first (left) and second (right) boundary-matching resonators. \textbf{d}, Transmission spectrum of the full resonator array consisting of $22$ unit cells and $2$ boundary-matching resonators on either end of the array (for a total of $26$ resonators).  Measured data is plotted as a red curve and the circuit model fit is plotted as a black curve. Fit model parameters are given in the text.}
\label{fig:taper}
\end{figure*}

\subsection{Geometrical Design of Unit Cell}
As shown in Fig.~\ref{fig:LinearDesign}, the unit cell of the resonator array in this work includes a lumped-element resonator formed from a tightly meandered wire with a large `head' capacitance, and `wing' capacitors which, in addition to providing the majority of the capacitance to ground, are used to couple between resonators in neighbouring unit cells.  The meandered wire has a $1$~$\mu$m pitch and a $1$~$\mu$m trace width for tight packing. From the top of the meander inductor is the head capacitor and a pair of thin metal capacitor strips which extend to the lateral edges of the unit cell (the wing capacitors).  The ground plane in between the resonators' meander inductor and the lateral wing capacitors acts as an electrical `fence', restricting the meander from coupling to neighboring resonators via stray capacitance or mutual inductance. This ensured that the bulk of the coupling between resonators was from the resonators' wing capacitive elements, thereby facilitating theoretical analysis of the structure using a simple single resonator per unit cell model. Furthermore, we included ground metal between the thin metal capacitor traces of neighbouring unit cell wing capacitors.  In this way, the ground planes above and below the resonator array are tied together at each unit cell boundary, thereby suppressing the influence of higher-order transverse, slot-line modes of the waveguide. 

In addition, anticipating integration with Xmon qubits, we incorporated into our unit cell design a Xmon shunting capacitance to ground, along with pads for facile addition of Josephson Junctions. This ensured that the addition of a qubit at a particular unit cell site in the resonator array minimally effected the capacitive environment surrounding that unit cell, and prevented the breaking of translational symmetry of the resonator array due to the addition of qubits. The capacitance between the Xmon capacitor and the rest of the unit cell was designed to be $\sim 2$~fF, yielding a qubit-unit cell coupling of $\gvacMMuc \approx 0.9J$.

\subsection{Matching of the Finite Resonator Array to Input-Output CPWs}

It has been previously shown that for a coupled cavity array, low-ripple transmission at the center of the passband is possible by appropriate variation of the inter-resonator coupling coefficients for a few of the resonators adjacent to the ports, effectively matching the finite periodic structure to the input-output ports~\cite{sumetsky2003modeling}. In the case of capacitively coupled electrical resonators, modifying the coupling capacitance in isolation results in a renormalization of the resonance frequency and thus constitutes a scattering center for propagating light. Thus, concurrent modification of both the coupling capacitance and the shunt capacitance to ground for the boundary resonators is necessary to achieve low-ripple transmission in the middle of the passband, as previously shown in filter design theory~\cite{cohn1957direct}. By constraining the total capacitance in each modified resonator to remain constant (and keeping the inductance constant), the total number of parameters to adjust in order to achieve low ripple transmission is merely equal to the chosen number of resonators to be modified, resulting in a low-dimensional optimization problem. A filter design software such as Microwave Office can be used to provide initial guesses on the optimal circuit parameters with high accuracy, which can then be further optimized.  

In the Main text we present results on matching of a resonator array spanning $26$ resonators to \fiftyOhm~CPWs via modification of two resonators at each of the array-CPW boundaries. The geometrical design of the boundary resonators are shown in Fig.~\ref{fig:taper}.  The number of boundary resonators to modify ($2$) was chosen as a compromise between device simplicity and spectral bandwidth over which matching occurs. In principle, however, more resonators could have been used for matching of the finite structure to the ports in order to decrease the ripples in transmission passband near the bandedges. Referring to the notation in Fig.~\ref{fig:taper}b, the targets for the boundary resonator elements extracted from Sonnet~\cite{sonnet} electromagnetic simulations of the unit cell, were $C_{1g} = 89$~fF, $C_{2g} = 8.9$~fF, $\coupcap = 6.47$~fF, $C_{1} = 269$~fF, $C_{2} = 351$~fF, $\gndcap = 353$~fF, and geometric inductance $L_0 = 2.92~\text{nH}$. The individual capacitive and inductive elements have parasitic inductance and capacitance, respectively, and thus were not simulated separately. Rather, circuit parameters for the three boundary resonators were extracted by simulating the whole unit cell. We extracted the circuit element parameters from these simulations by numerically obtaining the dispersion for an infinite array of each of the three types of resonator unit cells via the ABCD matrix method ~\cite{pozar2009microwave}.  This yielded $\omega_0$ and $\coupcap/\gndcap$; $\coupcap$ was obtained from the $B$ parameter of the $ABCD$ matrix (which contains information on the series impedance of the unit cell circuit). We found this method of extracting parameters from simulation to give much higher accuracy when compared to other approaches, such as simulating unit cell elements separately.

Figure~\ref{fig:taper}d shows a plot of the measured transmission spectrum of the fabricated $26$ unit cell slow-light waveguide based upon the above design and presented in the Main text (c.f., Fig.~\ref{fig:LinearDesign}). A circuit model fit to the measured transmission spectrum yields the following circuit element parameters for boundary and central waveguide unit cells: $C_{1g} = 87.5$~fF, $C_{2g} = 7.3$~fF, $\coupcap = 5.05$~fF, $C_{1} = 352.1$~fF, $C_{2} = 275.5$~fF, $\gndcap = 353.2$~fF, and geometric inductance $L_0 = 3.151\text{nH}$. Based upon this model fit, we were thus able to realize good correspondence (within 3\%) between design and measured capacitances to ground, while extracted coupling capacitances are systematically lower by approximately $1.5$~fF. We attribute the systematically smaller coupling to stray mutual inductance between neighboring meander inductors, which tends to lower the effective coupling impedance between the resonators. The slightly larger fit inductance compared to design is to be expected as the kinetic inductance of the meander trace was not included in simulation.  According to Ref.~\cite{gao2008physics}, for a $1$~$\mu$m trace width and $120$~nm thick aluminum wire, the expected increase in the total inductance due to kinetic inductance is approximately $5$\% of the geometric inductance, in reasonable correspondence to the measured value. 

\section{Disorder Analysis}
\label{App:Disorder}
Fluctuations in the bare resonance frequencies of the lumped-element resonators making up the metamaterial waveguide breaks the translational symmetry of the waveguide, and effectively leads to random scattering of traveling waves between different Bloch modes. This scattering results in an exponential reduction in the probability that a propagating photon traverses across the entire length of the waveguide.  Furthermore, if the strength of scattering is large relative to the photon hopping rate, Anderson localization of light occurs where photons are completely trapped within the waveguide\cite{wiersma1997localization}. Thus, the aforementioned strategy for constructing a slow-light waveguide from an array of weakly coupled resonators is at odds with the inherit presence of fabrication disorder in any practically realizable device. Therefore, a compromise must be struck between choosing an inter-resonator coupling low enough to provide significant delay, but high enough such that propagation through the metamaterial waveguide is not significantly compromised by resonator frequency disorder. 

\begin{figure}[tbp]
\centering
\includegraphics[width = \columnwidth]{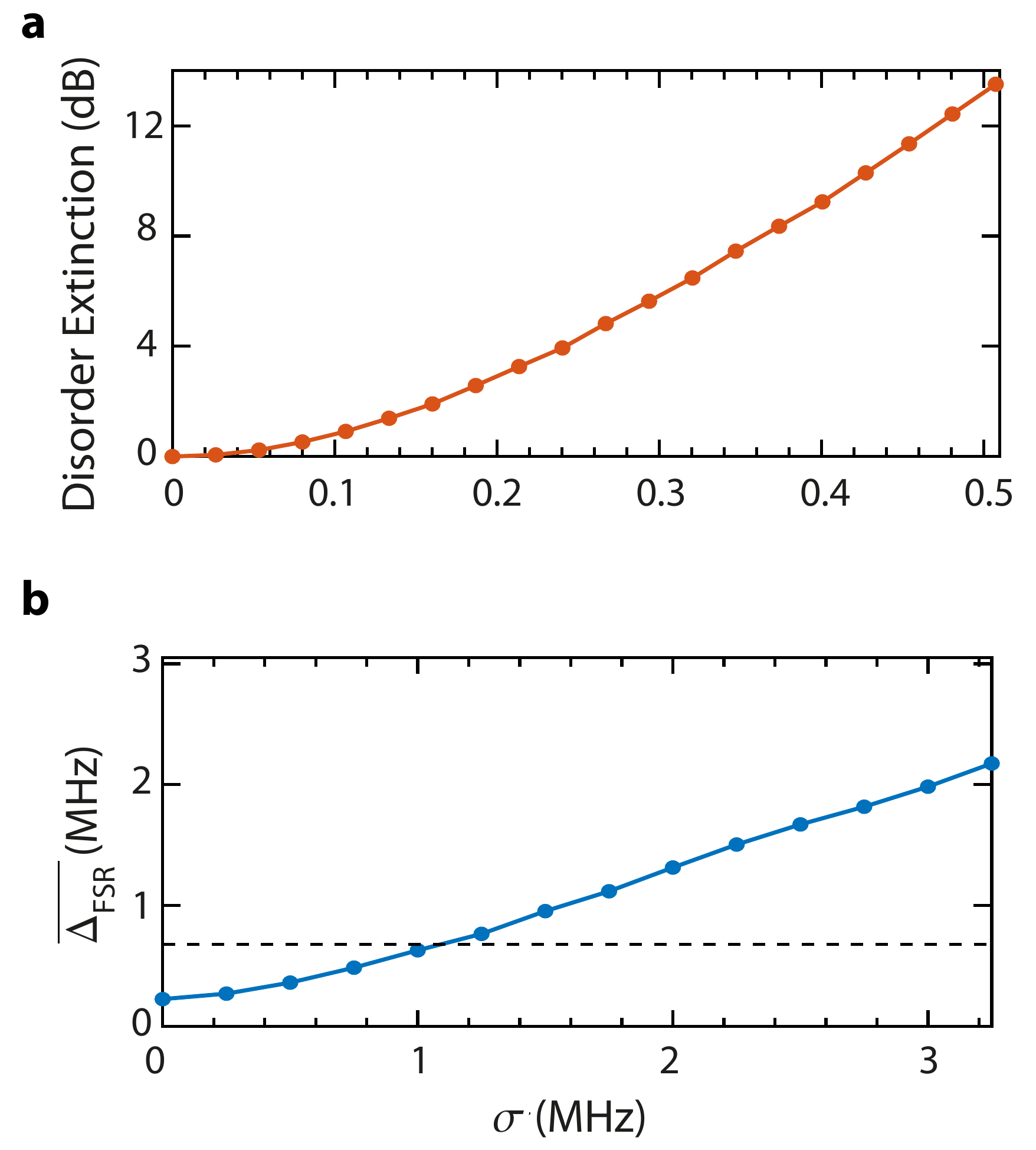}
\caption{\textbf{a}, Numerically calculated extinction as a function of disorder. Here, $\sigma$ is the disorder in the bare frequencies of the resonators making up the metamaterial waveguide and $J$ is the coupling between nearest-neighbour resonators in the resonator array. $50$ unit cells were used in this calculation, which included taper-matching sections at the input and output of the array that brought the overall passband ripple to 0.01dB. For a given disorder strength, $\sigma$, disorder extinction was calculated by taking the mean of the transmission across the passband for a given disorder realization, and subsequently averaging that mean transmission over many disorder realizations. Note that the calculated values depend on the number of unit cells. \textbf{b}, Numerically calculated variance in normal mode frequency spacing as a function of disorder. See text for details on the method of calculation of $\overline{\Delta_{\text{FSR}}}$. Dashed line indicates the experimentally measured $\Delta_{\text{FSR}}$, which was extracted from the data shown in Fig.~\ref{fig:LinearDesign}d.}
\label{fig:Disorder}
\end{figure}

Fig.~\ref{fig:Disorder}a shows numerical calculations of the transmission extinction in the metamaterial waveguide as a function of $\sigma/J$, where $\sigma$ is the resonator frequency disorder. This analysis was performed for a $50$ unit cell waveguide, with $\gndcap = 353.2$~fF, $\coupcap = 5.05$~fF, and $L_i = 3.101~\text{nH} + \delta_i$.  Here, $L_i$ is the inductance of the i$^{\text{th}}$ unit cell and $\delta_i$ are random inductance variations in each unit cell that give rise to a particular resonator frequency disorder, $\sigma$. These $L_i$ were calculated by: (i) determining the resonator frequencies of each unit cell by drawing from a Gaussian distribution with mean $\omegaMMres$ and variance $\sigma^2$, and (ii) solving for the corresponding inductances given the resonator frequencies and a fixed $\gndcap$. Note that we modeled the disorder as originating from inductance variations, rather than $\gndcap$ or $\coupcap$ variations, based on the fact that earlier work showed that disorder in superconducting microwave resonators was primarily due to variations in kinetic inductance~\cite{underwood2012low}. As we see in Fig.~\ref{fig:Disorder}a, in order for the average transmission to drop by less than 0.5dB ($10\%$), the normalized resonator frequency disorder must be less than $\sigma/J < 0.1$ .

In order to quantify the resonator frequency disorder in our fabricated resonator array one can analyze the passband ripple in transmission measurements~\cite{underwood2012low} (c.f., Fig.~\ref{fig:LinearDesign}d,e).  Given that the effect of tapering the circuit parameters at the boundary is to optimally couple the normal modes of the structure to the source and load impedances, the ripples in the passband are merely overlapping low-$Q$ resonances of the normal modes. Therefore, we can extract the normal mode frequencies from the maxima of the ripples in the passband, which will be shifted with respect the to normal mode frequencies of a structure without disorder. 

Furthermore, the mode spacing is dependent on the number of resonators and, in the absence of disorder, follows the dispersion relation shown in Fig.~\ref{fig:LinearDesign}c where the dispersion is relatively constant near the passband center and starts to shrink near the bandedges. In the presense of disorder, however, this pattern breaks down as the modes become randomly shifted. Our approach was therefore as follows.  Starting with the fit parameters presented in App.~\ref{App:Taper}, we simulated transmission through the metamaterial waveguide for varying amounts of resonator frequency disorder, $\sigma$.  For each level of disorder we performed simulations of $500$ different disorder realizations, and for each different disorder realization we computed the standard deviation in the free spectral range of the ripples, $\Delta_{\text{FSR}}$. This deviation in free spectral range was then averaged over all disorder realizations for each value of $\sigma$, yielding an empirical relation between $\bar{\Delta}_{\text{FSR}}(\sigma)$ and $\sigma$.  

\begin{figure*}[t]
\centering
\includegraphics[width = \textwidth]{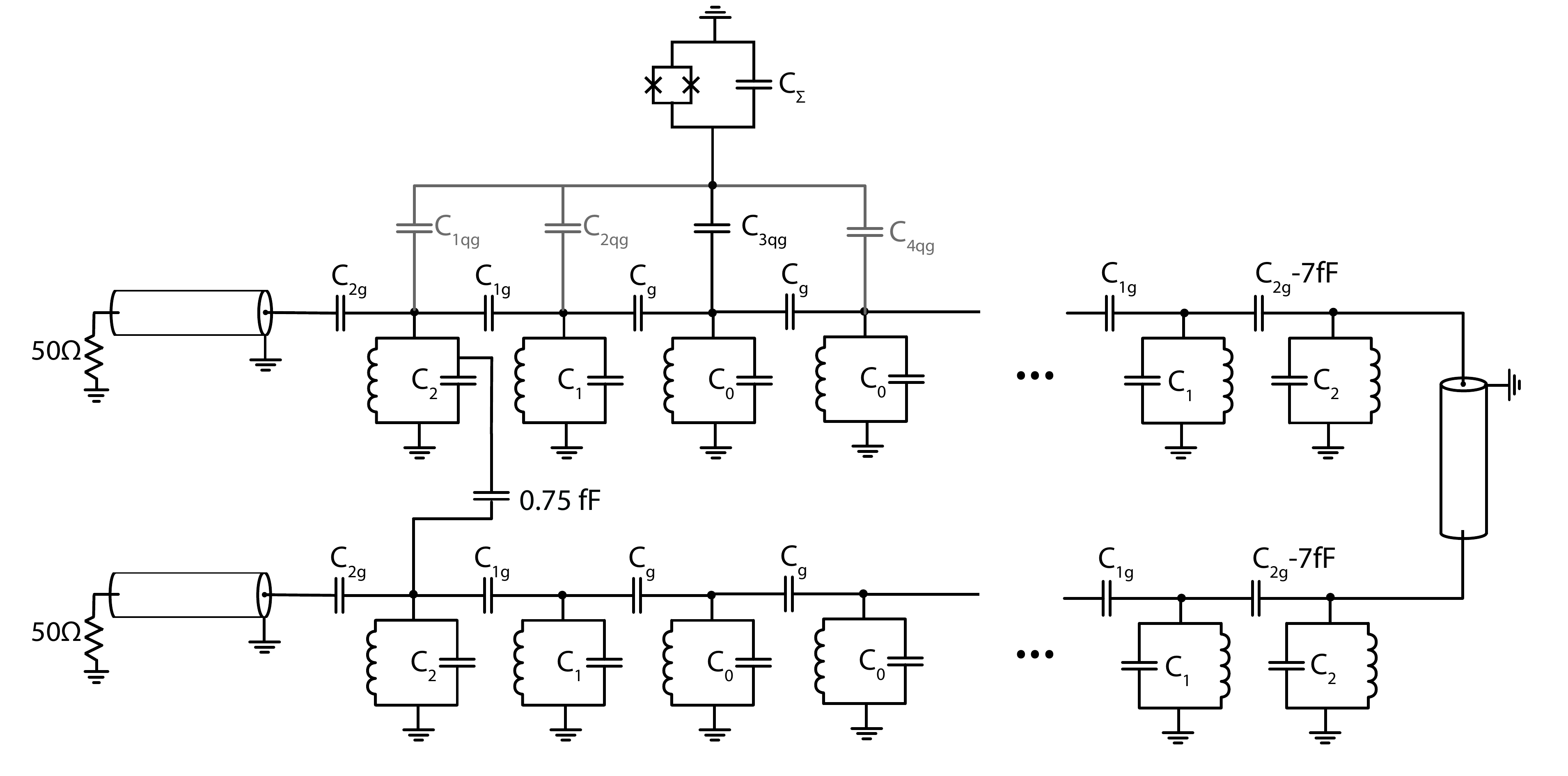}
\caption{Full circuit model used in simulations. All inductors were made equivalent, with inductance $L_0$.  Parameters are further discussed in the text.}
\label{fig:CircuitModel}
\end{figure*}

The numerically calculated empirical relation between variation in free spectral and frequency disorder is plotted in Fig.~\ref{fig:Disorder}b. Note that the minimum of $\bar{\Delta}_{\text{FSR}}(\sigma)$ at $\sigma=0$ is set by the intrinsic dispersion of the normal mode frequencies of the unperturbed resonator array.  As such, in order to yield a better sensitivity to disorder we chose to only use the center half of the passband in our analysis where dispersion is small.  From the data in Fig.~\ref{fig:LinearDesign}d, we calculated an experimental $\Delta_{\text{FSR}}$.  Comparing to the simulated plot of Fig.~\ref{fig:Disorder}b, this level of variance in the free spectral range results from a resonator frequency disorder within the array at the $1~\text{MHz}$ level (or $2\times 10^{-4}$ of the average resonator frequency), corresponding to $\sigma/J \approx 1/30$. We have extracted similar disorder values across a number of different metamaterial waveguide devices realized using our fabrication process.


\section{Modeling of Qubit Q$_{1}$ Coupled to the Metamaterial Waveguide}
\label{App:QubitModeling}

In this section we present modeling of the interaction between \Qone~and the metamaterial waveguide. We performed modeling via classical circuit analysis, where the qubit is represented by a linear resonator; this is an accurate representation of the qubit-waveguide system in the single-excitation limit. Time-resolved dynamical simulations were performed with the LTSpice numerical circuit simulation package, while frequency response simulations were performed with Microwave Office and standard circuit analysis. Our model, shown in Fig.~\ref{fig:CircuitModel}, assumes the following metamaterial waveguide parameters: $C_{2g} = 92.5$~fF, $C_{1g} = 7.8$~fF, $\coupcap = 5.02$~fF, $C_{2} = 273$~fF, $C_{1} = 351.2$~fF, $\gndcap = 353.2$~fF, and $L_0 = 3.099$~nH, which were obtained from fitting the transmission through the metamaterial device shown in Fig.~\ref{fig:ArraywithQubits}a with the qubit detuned away ($600$~MHz) from the upper bandedge. While in principle there are three independent parameters for every resonator (capacitance to ground, coupling capacitance, and inductance to ground), the set of metamaterial parameters above in addition to the qubit parameters were sufficient to achieve qualitative agreement between simulations and our data. 

Our model utilizes a qubit capacitance (excluding the capacitance to the metamaterial waveguide) of $C_\Sigma = 77.8$~fF, which, when assuming $E_c \approx -\hbar \annharm$, is consistent with measurements of the anharmonicity that was extracted by probing the two-photon transition between the $\ket{g}$ and $\ket{f}$ states. Furthermore, in the model we coupled the qubit to the first, third, and fourth resonators of the array (as opposed to just the third resonator), with capacitive couplings $C_{1qg}=0.16$~fF, $C_{3qg}=1.9$~fF, and $C_{4qg}=0.25$~fF. The coupling to resonators 1 and 4 was not intentional and was due to parasitic capacitance. We set $C_{2qg}=0$~fF in the model because the second metamaterial resonator was not expected to parasitically couple to the qubit as strongly as the first and fourth resonator due to the absence of an interdigitated capacitor and an integrated Xmon shunting capacitance (see Figure \ref{fig:taper} for images of the second resonator of the metamaterial waveguide). The  $C_{1qg}$ and $C_{4qg}$ parasitic capacitances were crucial to reproduce some of the subtle features in the measured data; this will be discussed in detail below. Finally, the $Q$-factor of each metamaterial resonator and the faux-qubit resonator was set to $9\times 10^4$ by incorporating losses in the inductors.

\subsection{Time Domain}

\begin{figure}[t]
\centering
\includegraphics[width = \columnwidth]{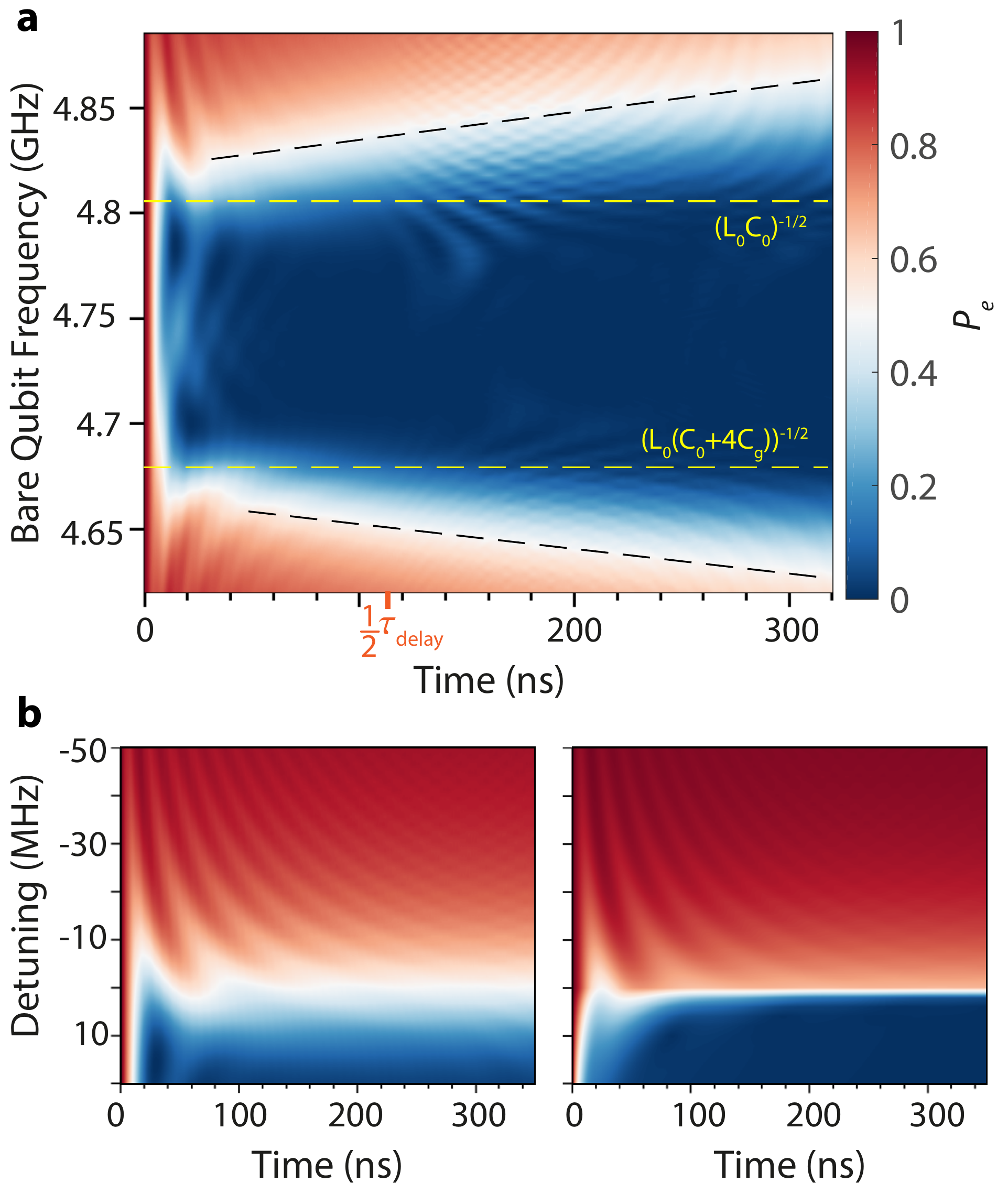}
\caption{\textbf{a}, Simulation of Fig.~\ref{fig:BandedgeDecay}b dataset. Simulation parameters are described in the text. Bandedges are highlighted in dashed yellow lines, while dashed black lines are guides to the eye. \textbf{b}, Comparison of the dynamics simulated by a modified circuit model of a qubit coupled to a metamaterial waveguide (left), and by population equations of motion derived in Ref.~\cite{john1994spontaneous} (right).  Refer to (\textbf{a}) for colorbar. Both models assume $\gvacMMuc/2\pi = 22~\text{MHz}$, corresponding to a reduction of the qubit's coupling to the waveguide by $30$\% as compared to the plot in (\textbf{a}), as well as $J/2\pi = 33.5~\text{MHz}$. See text for description of modified circuit model. We use $(\gvacMMuc^4/4J)^{1/3}$ in place of $\beta$ for simulations based on analysis in Ref.~\cite{john1994spontaneous}.}
\label{fig:SimTime}
\end{figure}

Figure~\ref{fig:SimTime}a shows the simulated dynamics of our circuit model as a function of bare qubit frequency (where the qubit inductance was swept to change the bare qubit frequency). It is evident that there is qualitative agreement between Fig.~\ref{fig:SimTime}a and the measured data in Fig.~\ref{fig:BandedgeDecay}b, indicating that our model captures the salient dynamical features of our measured data. The $C_{3qg}$ capacitance primarily sets the coupling of the qubit to the metamaterial waveguide. Its magnitude relative to $J$, along with the qubit frequency $\omegageint(\Phi)$, predominantly determines the frequency of oscillations near the bandedge, as well as the decay rate into the waveguide in the passband. In the absence of other parasitic capacitances, this decay rate is theoretically determined to be $\gvacMMuc^2 / v(\omegageint(\Phi))$~\cite{calajo2016atom}, where $v(\omegageint(\Phi))$ is the group velocity of the metamaterial waveguide at the qubit-waveguide interaction frequency $\omegageint(\Phi)$, and $\Phi$ is the applied flux through the SQUID loop used to tune the qubit frequency. 

The parasitic coupling $C_{4qg}$, however, is necessary to replicate the asymmetry in the dynamics near the upper and lower bandedges. This is because the lower bandedge modes have an oscillating charge distribution between unit cells, while the upper bandedge modes have a slowly-varying charge distribution across the unit cells. The parasitic coupling of the qubit to the neighboring unit cell therefore has the effect of lowering the qubit coupling to the lower bandedge modes due to cancellation-effects arising from the opposite charges on neighbouring resonators for lower bandedge modes. On the other hand, coupling of the qubit to the upper bandedge modes which have slowly-varying charge distributions, is enhanced. 

The $C_{1qg}$ coupling capacitance between the qubit and the resonator directly coupled to the \fiftyOhm~port is necessary to replicate the qubit population decay rate outside of the passband. This feature in the simulations and measured data, highlighted by dashed black lines in Fig.~\ref{fig:SimTime}a, is mostly due to loss from the finite overlap between the bound state and the external \fiftyOhm~environment of the input-output waveguides (which dominates the intrinsic loss due to \Qone's position near the boundary of the array), and would be flat for an infinite array. Note that, near the bandedges, it is this loss that results in a slow population decay as compared to the initial fast dynamics (see top panel of Fig.~\ref{fig:BandedgeDecay}c for a clear example). In the absence of the $C_{1qg}$ coupling, this overlap was not sufficiently high in the simulations given the coupling of the qubit to the metamaterial waveguide (extracted from separate measurements in the passband). Therefore, this overlap was made larger, while minimizing the increase to the overall coupling of the qubit to the metamaterial waveguide, by incorporating the small $C_{1qg}$ coupling to the first resonator of the array.

In addition, in simulations, the onset of oscillations seen at $\tau \approx 115~\text{ns}$ could be delayed or advanced by increasing or decreasing the number of resonators in between the qubit and the bend in the metamaterial waveguide model, while it could be removed altogether by removing the bend section. This indicated that these late time oscillations are a result of spurious reflection of the qubit's emission at the bend, due to the imperfect matching to the \fiftyOhm~coplanar waveguide in between the two resonator rows. Note that this impedance mismatch and reflections are amplified near the bandedges, where the Bloch impedance rapidly changes. Moreover, when the qubit frequency is near the bandedges, the reflected emission is distorted through its propagation in the metamaterial waveguide due to the significant dispersion near the bandedges. This results in a spatio-temporal broadening of the emitted radiation, which results in differences in both duration and amplitude between the early-time and late-time oscillations. The frequencies of both sets of oscillations, however, are set by $\gvacMMuc$ and $J$ as discussed in the main text. 

As alluded to in the main text, the early-time oscillations observed in our work are, qualitatively, a generic feature of the interaction between a qubit and a bandedge in a dispersive medium, and not merely an attribute of our specific system. In order to illustrate this point, in Fig.~\ref{fig:SimTime}b, we further compared the initial oscillations to the theory presented by John and Quang in Ref.~\cite{john1994spontaneous} of a qubit whose frequency lies in the spectral vicinity of a bandedge. For this comparison, we changed the circuit model of our system in the following manner: (i) we removed the parasitic coupling of the qubit to neighboring unit cells, in order to simplify the coupling to a single point coupling, (ii) we increased the size of the array to reduce boundary effects from the dynamics, (iii) we reduced the overall coupling of the qubit to the metamaterial waveguide, and (iv) we removed intrinsic losses. We undertook these changes in order to obtain better correspondence between our circuit model and the model assumed in Ref.~\cite{john1994spontaneous}, that of an atom (qubit) with point dipole coupling to an infinite periodic dielectric environment, whose frequency is in the spectral vicinity of only a single bandedge. Note, however, that the dispersion relation of the waveguide is different than the dispersion assumed in Ref.~\cite{john1994spontaneous}, and that persistent boundary effects remain. Nonetheless, above the bandedge, we see good qualitative agreement between the dynamics modeled both by the modified circuit model, and the population equation of motion derived in Ref.~\cite{john1994spontaneous}, with both simulations exhibiting very similar oscillatory decay to what is observed in Fig.~\ref{fig:SimTime}a and \ref{fig:BandedgeDecay}b. This further confirms our interpretation of the early-time non-Markovian dynamics in Fig.~\ref{fig:BandedgeDecay} discussed in the Main text, that the non-exponential oscillatory decay is due to the interaction between the qubit and the strong spike in the density of states at the bandedge. 

\subsection{Frequency Domain}

\begin{figure}[t]
\centering
\includegraphics[width = \columnwidth]{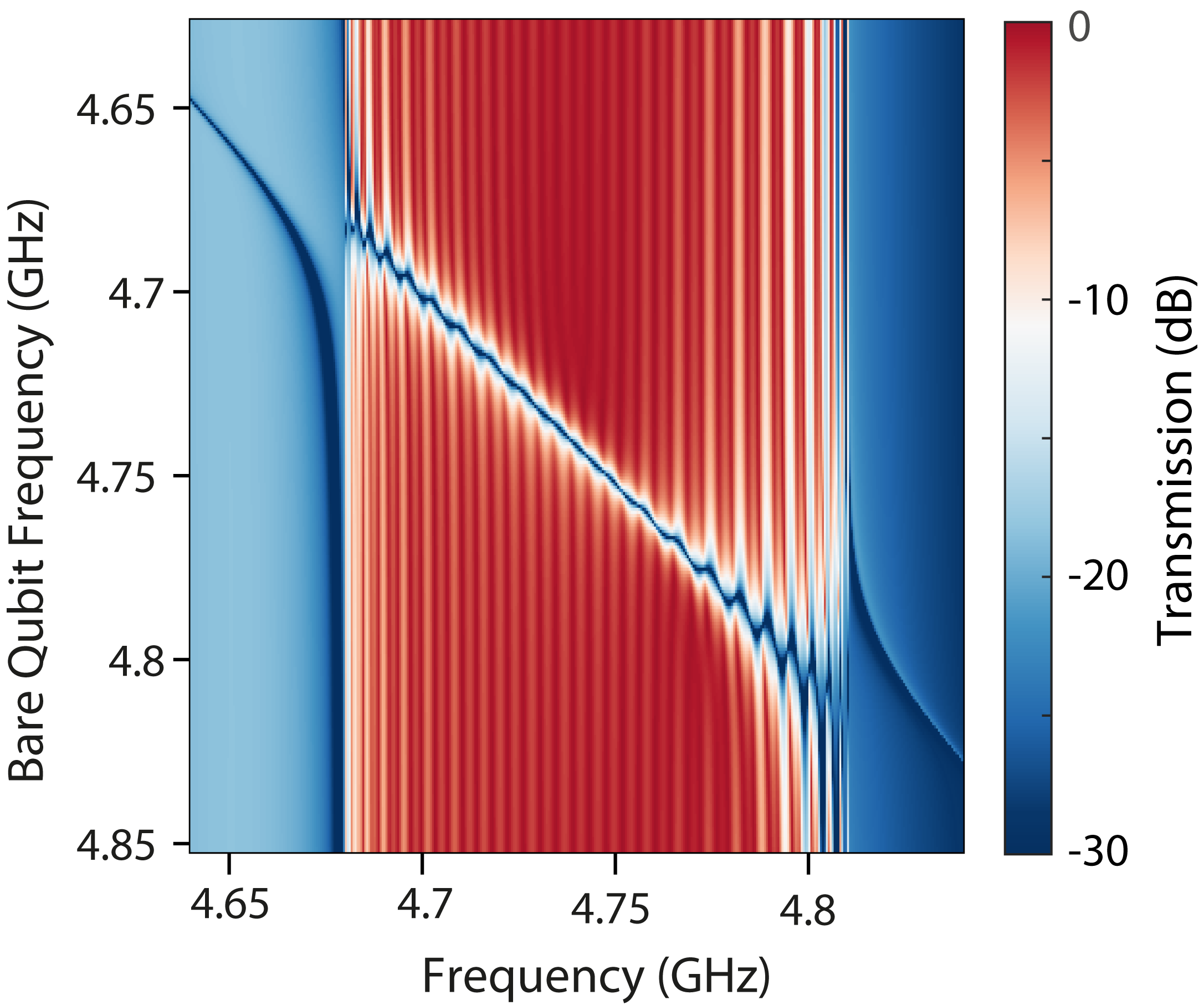}
\caption{Simulation of Fig.~\ref{fig:ArraywithQubits}b dataset. Circuit model and simulation parameters are described in the text. Simulations were done with the aid of the Microwave Office software package.}
\label{fig:SimFreq}
\end{figure}

As a final check of the accuracy of our circuit model in representing the fabricated qubit-waveguide system, in Fig.~\ref{fig:SimFreq} we plot an intensity color plot of the transmission through the slow-light waveguide as the bare qubit frequency is tuned across the passband using the circuit model (c.f., the corresponding measurement data plotted in Fig.~\ref{fig:ArraywithQubits}d). Note that in order to capture the background transmission levels as well as the interaction of the qubit with the background transmission, we included a small direct coupling capacitance of $0.75$~fF between the first and last resonators of the array. These two resonators have the largest crosstalk. This is due to the large portion of charge contained in the interdigitated capacitors between the resonators and the input-output waveguides.  In simulations without this background transmission, the qubit mode break-up near the bandedge, and signatures of the bound-state outside of the passband were significantly weaker. 

In addition, the series capacitance of the boundary resonators coupled to the input-output waveguides was made $7$~fF higher than the series capacitance of the boundary resonators coupled to the short CPW section in the bend, which is due to the proximity of the large bondpads used to probe the waveguides. Our simulations are in excellent qualitative agreement with the data presented in Fig.~\ref{fig:ArraywithQubits}d. They also captures the spectroscopic non-Markovian features of our data -- the repulsion of the bound state's energy from the bandedge and the persistence of the bound state even when the bare qubit frequency overlaps with the passband (see Refs.~\cite{calajo2016atom,sundaresan2019interacting,liu2017quantum} for further details). 

\vfill


%


\end{document}